\newif\ifonecolumn
\onecolumnfalse

\ifonecolumn
\documentclass[letterpaper,12pt,peerreviewca,draftcls]{IEEEtran}
\usepackage{csm16}
\else
\documentclass{IEEEtran}
\fi
\usepackage[margin=1in]{geometry}
\ifonecolumn
\usepackage[nolists,nomarkers,tablesfirst]{endfloat} %
\fi

\usepackage{amsmath} %

\IEEEoverridecommandlockouts                              %

\usepackage{url}
\usepackage{graphicx}
\usepackage{verbatim}%
\makeatletter
\newcommand{\verbatimfont}[1]{\def\verbatim@font{#1}}%
\makeatother
\usepackage{wrapfig}
\verbatimfont{\ttfamily\small}

\newcommand{\bi}{\begin{itemize}}\newcommand{\ei}{\end{itemize}}
\newcommand{\be}{\begin{equation}}\newcommand{\ee}{\end{equation}}
\newcommand{\bee}{\begin{enumerate}}\newcommand{\eee}{\end{enumerate}}
\newcommand{\bea}{\begin{eqnarray}}\newcommand{\eea}{\end{eqnarray}}
\newcommand{\beas}{\begin{eqnarray*}}\newcommand{\eeas}{\end{eqnarray*}}
\newcommand{\bc}{\begin{center}}\newcommand{\ec}{\end{center}}
\usepackage[left,pagewise]{lineno} 
\usepackage[english]{babel} 
\usepackage{blindtext}
\usepackage[compress]{cite}

\newif\ifhtml
\htmlfalse

\newif\ifPDF \ifx\pdfoutput\undefined\PDFfalse \else\ifnum\pdfoutput > 0\PDFtrue \else\PDFfalse \fi \fi
\ifPDF 
\usepackage[pdftex, plainpages = false, colorlinks=true, linkcolor=black, citecolor = green!50!blue, urlcolor = blue, filecolor=black, pagebackref=false, hypertexnames=false,  pdfpagelabels ]{hyperref}
\fi

\usepackage{amsmath}
\usepackage[T1]{fontenc}
\usepackage[latin9]{inputenc}
\usepackage{amsbsy}
\usepackage{amstext}
\usepackage{amssymb}
\usepackage{graphicx}
\usepackage{subcaption}
\ifonecolumn\else
\fi
\usepackage{color}
\usepackage{pdflscape}
\usepackage{pifont}
\usepackage{hyperref}
\usepackage{booktabs,multirow}
\usepackage[usenames,dvipsnames,svgnames,table]{xcolor}
\usepackage{tikz}
\usepackage{floatpag}

\newcommand{\coloredtwopiehole}[4]{%
\begin{tikzpicture}
 \draw[#3] (0,0) circle (.8ex);\fill[rotate=90-#2,fill=#3] (.8ex,0) arc (0:-#1:.8ex) -- (0,0) -- cycle;
  \draw[white] (0,0) circle (#4);\fill[rotate=90-#2,fill=white] (#4,0) arc (0:-#1:#4) -- (0,0) -- cycle;
\end{tikzpicture}%
}

\newcommand{\coloredtwopie}[3]{%
\begin{tikzpicture}
 \draw[#3] (0,0) circle (.8ex);\fill[rotate=90-#2,fill=#3] (.8ex,0) arc (0:-#1:.8ex) -- (0,0) -- cycle;
\end{tikzpicture}%
}

\newcommand{\coloredpie}[2]{%
\coloredtwopie{#1}{0}{#2}%
}

\newcommand{\pie}[2]{%
\coloredtwopie{#1}{0}{blue}%
}

\newcommand{\quarterpie}{\pie{90}~}
\newcommand{\halfpie}{\pie{180}~}
\newcommand{\threequarterpie}{\pie{270}~}
\newcommand{\fullpie}{\pie{360}~}
\newcommand{\piehole}{\coloredtwopiehole{360}{0}{blue}{0.3ex}~}

\newcommand{\emptycoloredpie}[1]{\coloredpie{0}{#1}~}

\newcommand{\halfcoloredpie}[1]{\coloredpie{180}{#1}~}
\newcommand{\threequartercoloredpie}[1]{\coloredpie{270}{#1}~}
\newcommand{\fullcoloredpie}[1]{\coloredpie{360}{#1}~}

\graphicspath{{./},{./fig/},{../},{../fig/}}
\makeatletter
\def\input@path{{./},{./sections/},{../},{../sections/}}
\makeatother

\usepackage{subfiles}

\newif\ifjournalv
\journalvtrue

\newif\ifmargincomments
\margincommentstrue

\ifmargincomments

\else

\fi

\newcommand{\revXXIa}[1]{#1}

\newcommand{\BCM}{\textcolor{blue}{\ding{51}} }

\title{Multi-Agent Algorithms for \\Collective Behavior\\
\Large A structural and application-focused atlas}
\author{Federico Rossi, Saptarshi Bandyopadhyay,
		Michael T. Wolf and Marco Pavone \\
		\ifonecolumn
		POC: F. Rossi (\href{mailto:federico.rossi@jpl.nasa.gov}{federico.rossi@jpl.nasa.gov} )
		\fi
}

\begin{document}

\maketitle
\ifonecolumn
\CSMsetup
\linenumbers \modulolinenumbers[2] %
\fi

The goal of this paper is to provide a survey and application-focused atlas of collective behavior coordination algorithms for multi-agent system, classified according to their underlying mathematical structure.

Multi-agent robotic systems hold promise to enable new classes of missions in space, aerial, terrestrial, and marine applications and deliver higher resilience and adaptability at lower cost compared to existing monolithic systems. Example applications include coordinated patrols of UAVs (uninhabited aerial vehicles), satellite formations for astronomy and Earth observation, and multi-robot planetary exploration. Many algorithms have been proposed to control the collective behavior of such systems, ranging from low-level position control to high-level motion planning and  task allocation algorithms.

Many excellent surveys of algorithms for collective behavior exist in the literature; however, such papers generally focus either on single applications -- e.g., formation control \cite{Ref:Oh15survey}, coverage \cite{Ref:Schwager09}, or task allocation \cite{Ref:Mataric04} -- or on specific control techniques -- e.g., consensus \cite{Ref:Garin10,Ref:Cao13} \revXXIa{and multi-agent reinforcement learning \cite{zhang2019MARL}}.
Several works study the fundamental limitations of performance of multi-agent systems: e.g., \cite{MartinezBulloEtAl2007a} and \cite{Ref:Rossi14} explore time and communication complexity in synchronous and asynchronous systems, respectively, and \cite{GuptaLangbortEtAl2006} studies robustness to agent failures.
However, these works only survey the performance of a limited number of applications and algorithms.

In contrast, in this paper we survey the \emph{general} family of collective behavior algorithms for multi-agent systems and classify them according to their underlying mathematical \emph{structure}, without  restricting our focus to specific applications or individual classes of algorithms. In doing so, we aim to capture fundamental mathematical properties of algorithms (e.g., scalability with respect to the number of agents and bandwidth use) and to show how the same algorithm or family of algorithms can be used for multiple tasks and applications.

\revXXIa{
Specifically, our contribution is threefold. 
First, we provide a classification of collective behavior algorithms according to their mathematical structure.
Second, we assess the applications that each class of algorithm is well-suited for, based on their use in the literature.
Third, we assess  the scalability, bandwidth requirements, and demonstrated maturity of algorithms in each class, based on their mathematical structure and on their use in the literature.

Collectively, these contributions provide an \emph{application-focused atlas} of algorithms for collective behavior  of multi-agent systems, with three objectives:
}
\begin{itemize}
\item to act as a tutorial guide to practitioners in the selection of coordination algorithms for a given application;
\item to highlight how mathematically similar algorithms can be used for a variety of tasks, ranging from low-level control to high-level coordination;
\item to explore the state-of-the-art in the field of control of multi-agent systems and identify areas for future research.
\end{itemize}

\ifhtml
\else
\subsection{Methodology}
\fi

\ifhtml
\else
\subsubsection{Literature Review} 
\fi

\revXXIa{
We performed a thorough and systematic literature search of major controls and robotics journals and conferences. Specifically, we searched the archives of the International Journal of Robotics Research, the Journal of Field Robotics, Autonomous Robots, and all IEEE journals indexed on IEEEXplore (including the Transactions on Robotics, the Transactions on Control of Networked Systems, the Control Systems Magazine, and the Robotics \& Automation Magazine) with a selection of keywords including, among others, "multi-robot", "multi-agent", "swarm", and "collective behavior". We strived to include in the survey all papers satisfying these queries that received more than 100 citations. We also included a number of less-cited journal papers and selected conference papers that we identified through references in the aforementioned papers and through the authors' background and expertise.
}

It is not feasible to cite \emph{all} existing works on multi-agent collective behavior algorithms; however, we have focused on identifying and classifying the key mathematical \emph{structures} and \emph{techniques} for coordination of multi-robot systems.

\begin{figure*}[t]
\begin{minipage}[c][\hsize][c]{\hsize}
\thisfloatpagestyle{empty}
\ifhtml
\else
\centering
\includegraphics[width=\textwidth]{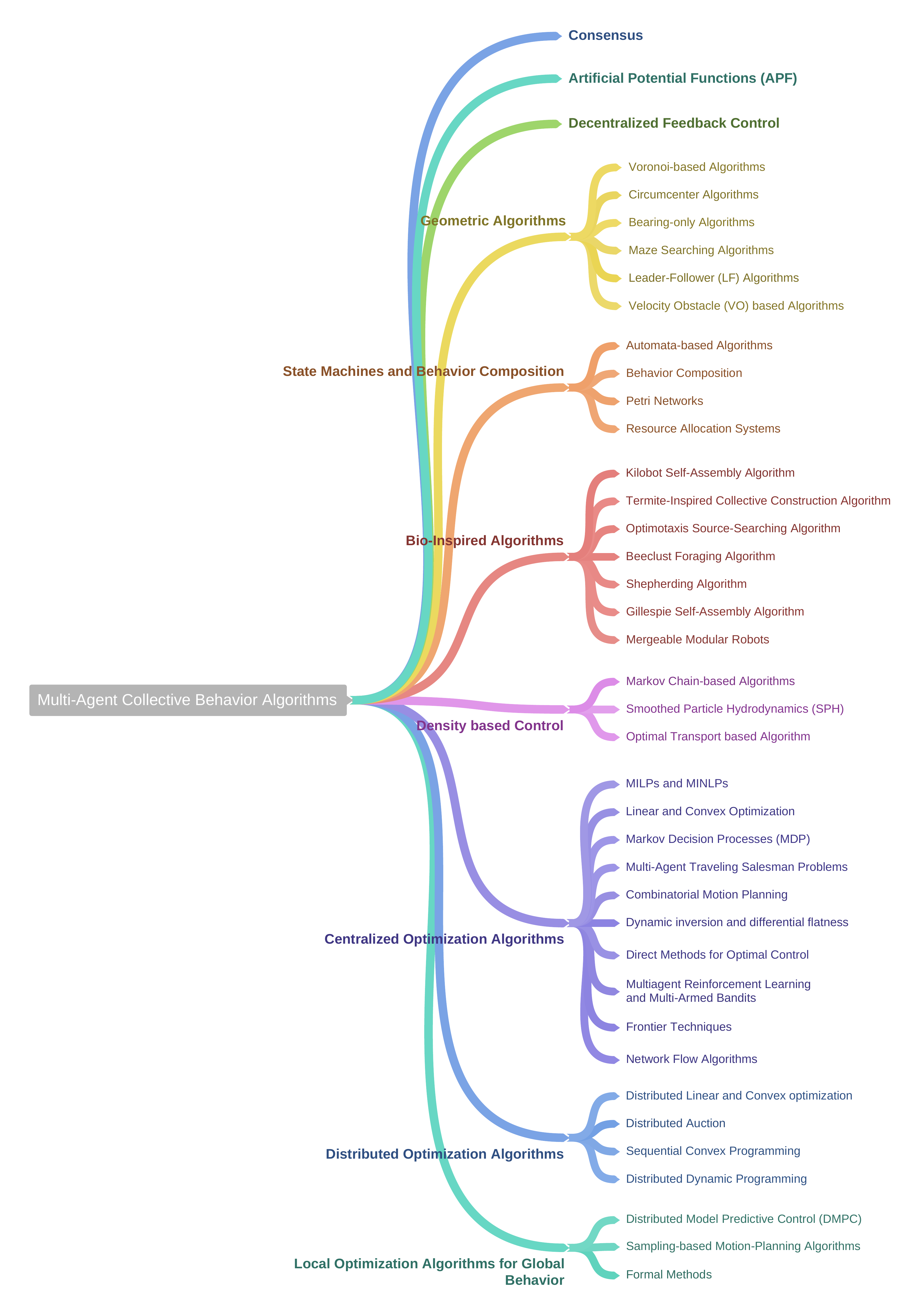}
\fi
\caption{Graphical representation of the classification of mathematical techniques for multi-agent collective behavior.\label{fig:all_algos}}
\end{minipage}
\end{figure*}

\begin{figure*}[t]
\begin{minipage}[c][\hsize][c]{\hsize}
\centering
\ifhtml
\else
\includegraphics[height=\textheight]{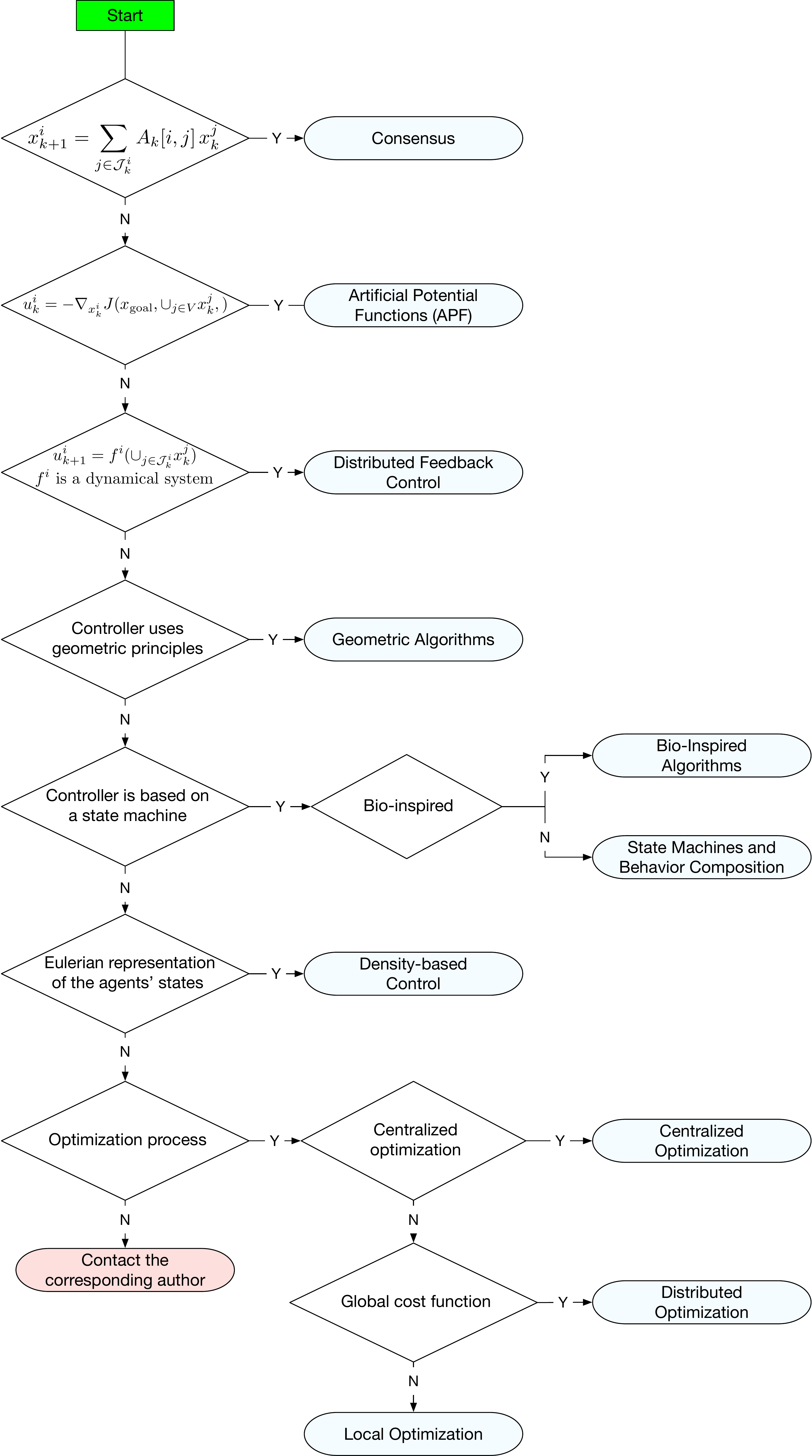}
\fi
\caption{Flow-chart used for classification of families of algorithms for collective behavior according to their mathematical structure.\label{fig:flowchart}}
\end{minipage}
\end{figure*}

\ifhtml
\else
\subsubsection{Classification} 
\fi
\revXXIa{
We grouped the surveyed algorithms in ten broad classes and a number of sub-classes according to their mathematical structure, as shown in Figure \ref{fig:all_algos}.
A synthetic description of each class is reported in the \nameref{sidebar:algorithm-classification} sidebar; a detailed description is presented in the remainder of this paper.

To perform the classification, we developed the flowchart shown in Figure \ref{fig:flowchart}. 
In devising the classification, we followed three principles:

\paragraph{Mathematical structure} We examined the mathematical underpinnings of each proposed algorithm, and identified commonalities in their structures. For instance, certain algorithms leverage feedback control based on the agents' states and their neighbors'; others rely on discrete state machines; and others rely on centralized or distributed optimization. 

\paragraph{Deference to existing categorizations in the literature} In compiling our classification, we followed existing categorizations in the literature, even when they apply different labels to algorithms with  similar mathematical structure. For instance, consensus algorithms are a special case of distributed feedback algorithms; nevertheless, we identified consensus as a standalone class due to its prominence in the robotics and control literature and to the wealth of theoretical results and applications developed specifically for consensus. As another example, the majority of bio-inspired algorithms rely on a state machine formulation; nevertheless, they are typically classified as distinct from state machines in the robotics literature, a categorization that we followed in this work.

\paragraph{Authors' experience} The proposed classification is  based on the authors' experience, and it is necessarily opinionated. We recognize that the proposed classification is not the only possible one and, in particular, we do not claim that this classification is a rigorous \emph{taxonomy} of algorithms for collective behavior. Nevertheless, we hope that the classification can serve as an \emph{atlas} of algorithms for collective behavior, guiding researchers and practitioners through the vast literature, identifying commonalities between algorithmic approaches, and mapping classes of algorithms that are well-suited for a given application.
}

\ifhtml
\else
\subsubsection{Figures of Merit} 
\fi

We analyzed the scalability, bandwidth use, and demonstrated maturity of each mathematical technique according to the following metrics.
\revXXIa{
The symbols shown in the text are used to identify each performance class in Table \ref{tab:all_algos}.
}

\paragraph{Scalability} we classify mathematical techniques that perform well on systems with up to 5 agents \quarterpie, 20 agents \halfpie, 100 agents \threequarterpie, and over 1000 agents \fullpie. Certain techniques (e.g., Eulerian approaches) are only applicable to systems with large numbers of agents, i.e. more than 100 agents \piehole.

\paragraph{Bandwidth use} we assume that agents employ a point-to-point communication mechanism. That is, the bandwidth required to relay the same message to $k$ agents in a given time window is $k$ times the bandwidth required to relay the message to a single agent. We focus our analysis on the rate of growth of bandwidth use with the number of agents. Specifically, we identify four classes of algorithms: \\
\fullcoloredpie{blue} ~\emph{zero-communication} algorithms require no inter-agent communication; coordination is achieved exclusively through sensory information;\\ %
\threequartercoloredpie{blue} ~in \emph{low-bandwidth} algorithms, each agent communicates with a single neighbor at each time step; examples include gossip algorithms, leader-follower architectures, and algorithms that rely on a tree communication topology;\\ %
\halfcoloredpie{blue} in ~\emph{medium-bandwidth} algorithms, agents communicate with all their neighbors at each time step;\\ %
\emptycoloredpie{blue} ~\emph{high-bandwidth} algorithms require every agent to communicate with all other agents (including non-neighbors), potentially through multiple communication hops, at each time step.
This class of algorithms includes \emph{centralized} algorithms, which can be implemented with a shared-world communication architecture that continuously synchronizes relevant information across all agents.

\paragraph{Maturity} The three classes of algorithms are: \\
\emptycoloredpie{RedOrange}~ only demonstrated in ``simulation''; \\
\halfcoloredpie{ForestGreen}~ demonstrated in ``hardware''  either in the lab or in technology demonstration missions;\\
\fullcoloredpie{Blue}~ demonstrated in ``field'' deployments as part of a larger coordination architecture (excluding technology demonstrator missions).

\ifhtml
\else
\subsubsection{Applications of algorithms for collective behavior} 
\fi

\revXXIa{
For each class of algorithms, we identified applications well-suited for it based on usage in the literature. To do so, we identified nine classes of applications of interest, closely following the taxonomy of tasks in multi-agent systems proposed in \cite{Ref:Brambilla13}.
}

\begin{enumerate}
\item \textbf{Spatially-organizing behaviors}, where agents coordinate to achieve a given spatial configuration and have negligible interactions with the environment. These tasks can be further classified into:
\begin{enumerate}
\item \textit{Aggregation:} converging to one location.
\item \textit{Pattern Formation:} achieving a desired formation.
\item \textit{Coverage:} covering an area.
\end{enumerate}
\item \textbf{Collective explorations}, where agents interact with the environment but have minimal interaction among themselves. These tasks can be classified into:
\begin{enumerate}
\item \textit{Area Exploration:} exploring the environment for mapping or surveillance.
\item \textit{Goal Searching:} searching for targets. 
\end{enumerate}
\item \textbf{Cooperative decision making}, where agents both coordinate among themselves and interact with the environment to accomplish complex tasks. These tasks can be further classified into:
\begin{enumerate}
\item \textit{Task Allocation:} distributing tasks among agents.
\item\textit{Collective Transport:} coordinating to transport large objects.
\item \textit{Motion Planning:} finding paths or trajectories in cluttered environments.
\item \textit{Distributed Estimation:} estimating the state of one or multiple targets.
\end{enumerate}
\end{enumerate}
These simple tasks are the fundamental building blocks of many complex multi-agent applications.

\ifhtml
\else
\subsection{Organization} 
\fi

Our key contribution is Table \ref{tab:all_algos}, which shows the proposed classification of mathematical techniques for collective behavior, highlights the tasks that each mathematical technique is well-suited for, and reports figures of merit. A detailed graphical representation of the proposed atlas is also presented in Figure \ref{fig:all_algos}.

In the remainder of this paper we provide a brief description of the identified mathematical techniques for collective behavior, a succinct discussion of the mathematical formulation and analytical guarantees of selected techniques,  and relevant references.  Finally, we conclude the paper with some inferences and suggest directions for future research in the \emph{Conclusions} section.

\revXXIa{
A preliminary version of this paper was presented at the 2018 IFAC Aerospace Controls TC Workshop on Networked \& Autonomous Air \& Space Systems \cite{RossiBandyopadhyayEtAl2018}. In this revised and extended version, we provide a significantly expanded  discussion of each class of algorithms, an updated classification of algorithms in classes, and an in-depth description of the methodology followed in compiling the atlas.  
}

\ifhtml
\else
\subsection{Notation} 
\fi
We adopt a discrete-time framework. 
The communication network topology at the $k^{\text{th}}$ time step is represent by the graph $\mathcal{G}_{k}=(\mathcal{V},\mathcal{E}_{k})$, where $\mathcal{V}$ represents the set of agents and $\mathcal{E}_{k}$ represents the time-varying set of directed edges in the graph.
If the $i^{\text{th}}$ agent receives information from the $j^{\text{th}}$ agent at the $k^{\text{th}}$ time step, then  edge $\overrightarrow{ji}\in\mathcal{E}_{k}$.
The set of neighbors that the $i^{\text{th}}$ agent receives information from at the $k^{\text{th}}$ time step are denoted by $\mathcal{N}_{k}^{i}$. The set of inclusive neighbors of the $i^{\text{th}}$ agent at
the $k^{\text{th}}$ time step are denoted by $\mathcal{J}_{k}^{i}=\mathcal{N}_{k}^{i}\cup\{i\}$. 

Let $\mathbb{N}$ and $\mathbb{R}$ be the set of natural numbers (positive integers) and real numbers, respectively.
Let $x_{k}^{i}\in\mathbb{R}^{n_{x}}$ represent the state of the $i^{\text{th}}$ agent at the the $k^{\text{th}}$ time step, where $n_{x}$ represents the dimension of the state.

\clearpage

\clearpage

\thispagestyle{empty}
\clearpage

{
\newcommand{\rot}[1]{\multicolumn{1}{c}{\rotatebox{90}{#1}}}

\ifhtml
\else
\rowcolors{2}{gray!25}{white}
\fi
\begin{table*}[p]
\thisfloatpagestyle{empty}
\begin{minipage}[t][\textheight][c]{\hsize}
\centering
\small
\makebox[\textwidth]{
\begin{tabular}{l|lll|ll|llll||lll}
\ifhtml
\else
\toprule
\fi
{} & \rotatebox{90}{Aggregation} & \rotatebox{90}{Pattern Formation} & \rotatebox{90}{Coverage} & \rotatebox{90}{Area Exploration} & \rotatebox{90}{Goal Searching} & \rotatebox{90}{Task Allocation} & \rotatebox{90}{Collective Transport} & \rotatebox{90}{Motion Planning} & \rotatebox{90}{Distributed Estimation} & \rotatebox{90}{Scalability} &  \rotatebox{90}{Bandwidth Use} &      \rotatebox{90}{Maturity} \\
 \midrule
 \hline \hyperref[sec:Consensus-algorithm]{\textbf{Consensus}}                                             &  \BCM &  \BCM &  \BCM &   &   &   &   &   &  \BCM &  \fullpie &  \halfcoloredpie{blue} &  \halfcoloredpie{ForestGreen} \\
 \hline \hyperref[sec:potential-functions]{\textbf{Artificial Potential Functions (APF)}}                  &  \BCM &  \BCM &  \BCM &  \BCM &   &  \BCM &  \BCM &  \BCM &   &  \fullpie &  \halfcoloredpie{blue} &  \fullcoloredpie{blue} \\
 \hline \hyperref[sec:distributed-feedback]{\textbf{Distributed Feedback Control}}                         &  \BCM &  \BCM &   &   &   &   &   &   &  \BCM &  \fullpie &  \halfcoloredpie{blue} &  \fullcoloredpie{blue} \\
 \hline \hyperref[sec:geometric-algo]{\textbf{Geometric Algorithms}}                                       &   &   &   &   &   &   &   &   &   &   &   &   \\
 \hyperref[subsec:Voronoi]{Voronoi-based Algorithms}                                                       &  \BCM &   &  \BCM &  \BCM &   &   &   &  \BCM &   &  \fullpie &  \halfcoloredpie{blue} &  \halfcoloredpie{ForestGreen} \\
 \hyperref[subsec:circumcenter]{Circumcenter Algorithms}                                                   &  \BCM &  \BCM &   &   &   &   &   &   &   &  \fullpie &  \halfcoloredpie{blue} &  \emptycoloredpie{RedOrange} \\
 \hyperref[subsec:bearing-only]{Bearing-only Algorithms}                                                   &  \BCM &  \BCM &   &   &   &   &   &   &   &  \fullpie &  \halfcoloredpie{blue} &  \halfcoloredpie{ForestGreen} \\
 \hyperref[subsec:maze-searching]{Maze Searching Algorithms}                                               &   &   &   &   &   &   &   &  \BCM &   &  \quarterpie &  \fullcoloredpie{blue} &  \emptycoloredpie{RedOrange} \\
 \hyperref[subsec:leader-follower]{Leader-Follower (LF) Algorithms}                                        &   &  \BCM &   &   &   &   &   &   &   &  \fullpie &  \threequartercoloredpie{blue} &  \emptycoloredpie{RedOrange} \\
 \hyperref[subsec:velocity-obstacle]{Velocity Obstacle (VO) based Algorithms}                              &   &   &   &   &   &   &   &  \BCM &   &  \fullpie &  \fullcoloredpie{blue} &  \fullcoloredpie{blue} \\
 \hline \hyperref[sec:automata-algo]{\textbf{State Machines and Behavior Composition}}                     &   &   &   &   &   &   &   &   &   &   &   &   \\
 \hyperref[subsec:message-passing]{Automata-based Algorithms}                                              &   &   &   &   &   &  \BCM &   &   &  \BCM &  \fullpie &  \threequartercoloredpie{blue} &  \emptycoloredpie{RedOrange} \\
 \hyperref[subsec:behavior-composition]{Behavior Composition}                                              &   &   &   &   &   &  \BCM &  \BCM &   &   &  \quarterpie &  \emptycoloredpie{blue} &  \halfcoloredpie{ForestGreen} \\
 \hyperref[subsec:petri-nets]{Petri Networks}                                                              &   &   &   &   &   &  \BCM &   &   &   &  \threequarterpie &  \emptycoloredpie{blue} &  \halfcoloredpie{ForestGreen} \\
 \hyperref[subsec:RAS]{Resource Allocation Systems}                                                        &   &   &   &   &   &   &   &  \BCM &   &  \quarterpie &  \emptycoloredpie{blue} &  \emptycoloredpie{RedOrange} \\
 \hline \hyperref[sec:bio-inspired-algo]{\textbf{Bio-Inspired Algorithms}}                                 &   &   &   &   &   &   &   &   &   &   &   &   \\
 \hyperref[subsec:kilobot]{Kilobot Self-Assembly Algorithm}                                                &   &  \BCM &   &   &   &   &   &   &   &  \piehole &  \halfcoloredpie{blue} &  \halfcoloredpie{ForestGreen} \\
 \hyperref[subsec:stigmergy]{Termite-Inspired Collective Construction Algorithm}                            &   &   &   &   &   &  \BCM &  \BCM &   &   &  \fullpie &  \fullcoloredpie{blue} &  \halfcoloredpie{ForestGreen} \\
 \hyperref[subsec:optimotaxis]{Optimotaxis and Immune-Inspired Source-Searching Algorithm}                 &   &   &   &   &  \BCM &   &   &   &   &  \piehole &  \fullcoloredpie{blue} &  \emptycoloredpie{RedOrange} \\
 \hyperref[subsec:beeclust]{Beeclust Foraging Algorithm}                                                   &   &   &   &  \BCM &   &   &   &   &   &  \piehole &  \fullcoloredpie{blue} &  \emptycoloredpie{RedOrange} \\
 \hyperref[subsec:shepherding]{Shepherding Algorithm}                                                      &  \BCM &   &   &   &  \BCM &   &   &   &   &  \fullpie &  \fullcoloredpie{blue} &  \emptycoloredpie{RedOrange} \\
 \hyperref[subsec:gillespie]{Gillespie Self-Assembly Algorithm}                                            &   &  \BCM &   &   &   &   &   &   &   &  \threequarterpie &  \halfcoloredpie{blue} &  \halfcoloredpie{ForestGreen} \\
 \hyperref[subsec:MNS]{Mergeable Modular Robots}                                                           &   &  \BCM &   &   &   &   &   &   &   &  \halfpie &  \threequartercoloredpie{blue} &  \halfcoloredpie{ForestGreen} \\
 \hline \hyperref[sec:density-based-control]{\textbf{Density-based Control}}                               &   &   &   &   &   &   &   &   &   &   &   &   \\
 \hyperref[subsec:markov-chain]{Markov Chain-based Algorithms}                                             &   &  \BCM &  \BCM &   &   &  \BCM &   &   &   &  \piehole &  \fullcoloredpie{blue} &  \halfcoloredpie{ForestGreen} \\
 \hyperref[subsec:SPH]{Smoothed Particle Hydrodynamics (SPH)}                                              &   &  \BCM &  \BCM &   &   &   &   &   &   &  \threequarterpie &  \halfcoloredpie{blue} &  \halfcoloredpie{ForestGreen} \\
 \hyperref[subsec:optimal-transport]{Optimal Transport based Algorithm}                                    &   &  \BCM &   &   &   &   &   &  \BCM &   &  \piehole &  \halfcoloredpie{blue} &  \emptycoloredpie{RedOrange} \\
 \hline \hyperref[sec:centralized-optimization-algo]{\textbf{Centralized Optimization Algorithms}}         &   &   &   &   &   &   &   &   &   &   &   &   \\
 \hyperref[subsec:MILP]{MILPs and MINLPs}                                                                  &   &  \BCM &   &   &   &  \BCM &   &  \BCM &  \BCM &  \quarterpie &  \emptycoloredpie{blue} &  \halfcoloredpie{ForestGreen} \\
 \hyperref[subsec:centralized-LP]{Linear and Convex Optimization}                                          &   &   &   &   &   &  \BCM &   &  \BCM &  \BCM &  \fullpie &  \emptycoloredpie{blue} &  \emptycoloredpie{RedOrange} \\
 \hyperref[subsec:MDP]{Markov Decision Processes (MDP)}                                                    &   &   &   &   &   &  \BCM &   &  \BCM &   &  \quarterpie &  \emptycoloredpie{blue} &  \halfcoloredpie{ForestGreen} \\
 \hyperref[subsec:TSP]{Multi-Agent Traveling Salesman Problems}                                            &   &   &  \BCM &  \BCM &  \BCM &  \BCM &   &   &   &  \quarterpie &  \emptycoloredpie{blue} &  \halfcoloredpie{ForestGreen} \\
 \hyperref[subsec:combinatorial-mp]{Coordinated Motion-Planning Algorithms}                                &   &   &   &   &   &   &   &  \BCM &   &  \halfpie &  \emptycoloredpie{blue} &  \emptycoloredpie{RedOrange} \\
 \hyperref[subsec:differential-flatness]{Dynamic inversion and differential flatness}                      &   &   &   &   &   &   &  \BCM &   &   &  \quarterpie &  \emptycoloredpie{blue} &  \halfcoloredpie{ForestGreen} \\
 \hyperref[subsec:direct-methods]{Direct Methods for Optimal Control}                                      &   &   &  \BCM &  \BCM &   &   &   &  \BCM &   &  \halfpie &  \emptycoloredpie{blue} &  \fullcoloredpie{blue} \\
 \hyperref[subsec:RL]{Multi-agent Reinforcement Learning}                                                  &   &   &   &  \BCM &  \BCM &  \BCM &   &   &  \BCM &  \quarterpie &  \emptycoloredpie{blue} &  \halfcoloredpie{ForestGreen} \\
 \hyperref[subsec:MAB]{\quad \emph{Multi-Armed Bandits}}                                                   &   &   &   &  \BCM &  \BCM &  \BCM &   &   &  \BCM &  \threequarterpie &  \emptycoloredpie{blue} &  \emptycoloredpie{RedOrange} \\
 \hyperref[subsec:centralized-frontier]{Frontier Techniques}                                               &   &   &   &  \BCM &  \BCM &   &   &   &   &  \halfpie &  \emptycoloredpie{blue} &  \fullcoloredpie{blue} \\
 \hyperref[subsec:network-flow]{Network Flow Algorithms}                                                   &   &   &   &   &   &  \BCM &   &  \BCM &   &  \fullpie &  \emptycoloredpie{blue} &  \emptycoloredpie{RedOrange} \\
 \hline \hyperref[sec:distributed-optimization-algorithms]{\textbf{Distributed Optimization Algorithms}}   &   &   &   &   &   &   &   &   &   &   &   &   \\
 \hyperref[subsec:LP]{Distributed Linear and Convex optimization}                                          &   &  \BCM &   &   &   &  \BCM &   &   &  \BCM &  \fullpie &  \halfcoloredpie{blue} &  \emptycoloredpie{RedOrange} \\
 \hyperref[subsec:auction]{Distributed Auction}                                                            &   &   &   &   &   &  \BCM &   &   &   &  \fullpie &  \emptycoloredpie{blue} &  \halfcoloredpie{ForestGreen} \\
 \hyperref[subsec:SCP]{Distributed Sequential Convex Programming}                                          &   &   &   &   &   &   &   &  \BCM &   &  \threequarterpie &  \halfcoloredpie{blue} &  \halfcoloredpie{ForestGreen} \\
 \hyperref[subsec:distributed-dp]{Distributed Dynamic Programming}                                         &   &   &   &   &   &  \BCM &   &  \BCM &   &  \halfpie &  \emptycoloredpie{blue} &  \halfcoloredpie{ForestGreen} \\
 \hline \hyperref[sec:local-optimization-algo]{\textbf{Local Optimization Algorithms for Global Behavior}} &   &   &   &   &   &   &   &   &   &   &   &   \\
 \hyperref[subsec:MPC]{Distributed Model Predictive Control (DMPC) }                                       &   &  \BCM &   &   &   &   &   &  \BCM &   &  \halfpie &  \halfcoloredpie{blue} &  \halfcoloredpie{ForestGreen} \\
 \hyperref[subsec:sampling-based-mp]{Decoupled and Prioritized Motion Planning Algorithms }                &   &   &   &   &   &   &   &  \BCM &   &  \fullpie &  \threequartercoloredpie{blue} &  \halfcoloredpie{ForestGreen} \\
 \hyperref[subsec:formal-methods]{Formal Methods}                                                          &   &   &   &   &   &   &   &  \BCM &   &  \fullpie &  \fullcoloredpie{blue} &  \emptycoloredpie{RedOrange} \\
 \bottomrule
 \end{tabular}
}
\caption{Categorization of collective behavior algorithms according to their mathematical structure and applicability of each algorithm to common multi-agent tasks. The scalability, bandwidth use, and level of demonstrated maturity of each algorithm are also reported. }\label{tab:all_algos}%
\end{minipage}
\end{table*}
}

\ifonecolumn\else
\clearpage
\fi

\section{Consensus Algorithm}
\label{sec:Consensus-algorithm}
\begin{figure}[h]
\centering
\includegraphics[width=.35\textwidth]{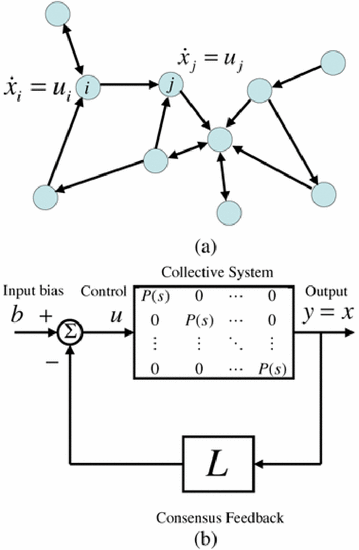}
\caption{Two equivalent forms of \textbf{consensus} algorithms: (a) a network of integrator agents in which agent $i$ receives the state $x_j$ of its neighbor, agent $j$, if there is a link $(i,j)$ connecting the two nodes; and (b) the block diagram for a network of interconnected dynamic systems all with identical transfer functions $P(s)=1/s$. The collective networked system has a diagonal transfer function and is a multiple-input multiple-output (MIMO) linear system.
In consensus algorithms, each agent updates its own state as a weighted average of its neighbors' states. The set of neighbors can change in time based on the agents' motion.
\copyright 2007 IEEE. Reprinted, with permission, from \cite{Ref:Saber07}.}
\label{fig:consensus:olfatisaber_murray}
\end{figure}

Consensus is among the oldest and most widely used techniques for coordination of multi-agent systems \cite{Ref:Tsitsiklis86,Ref:Jadbabaie03,Ref:Saber04,Ref:Ren05TAC}.
\revXXIa{
In consensus algorithms, agents repeatedly update their state as the average of their neighbors' states; the set of neighbors may change in time according to their motion.
Figure \ref{fig:consensus:olfatisaber_murray} (from \cite{Ref:Saber07}) shows a prototypical form of the consensus algorithm.
}

\textbf{\textit{Mathematical description:}} In the most general form of the consensus algorithm, each agent updates its state
using the following agreement equation:
\begin{equation}
x_{k+1}^{i}=\sum_{j\in\mathcal{J}_{k}^{i}}A_{k}[i,j]\thinspace x_{k}^{j}\thinspace,\quad\forall i\in\mathcal{V}\thinspace,\thinspace\forall k\in\mathbb{N}\thinspace,
\end{equation}
where the non-negative matrix $A_{k}\in\mathbb{R}^{\left|\mathcal{V}\right|\times\left|\mathcal{V}\right|}$ conforms to the graph $\mathcal{G}_{k}$ and is row stochastic, i.e., $\sum_{j\in\mathcal{V}}A_{k}[i,j]=1$ for all $i\in\mathcal{V}$.
If the graph $\mathcal{G}_{k}$ is  periodically-connected, then the states of all the agents converge to a unique state. 
If the matrix $A_{k}$ is balanced (i.e., doubly stochastic), then the final state is the average of all the agents' states (i.e., $\frac{1}{\left|\mathcal{V}\right|}\sum_{i\in\mathcal{V}}x_{1}^{i}$). 
Alternatively, if one of the agents (denoted as a leader) does not execute the consensus protocol but moves according to an exogenous trajectory, the other agents converge to the leader's state \cite{Ref:Hong06tracking,Ref:Ren2007ref}.
In \textit{gossip algorithms} \cite{Ref:Shah06}, each agent communicates with a single randomly-selected neighbor at each time step, minimizing bandwidth use.
In \textit{cyclic pursuit algorithms} \cite{Ref:Marshall04}, the consensus algorithm is executed on a directed ring communication topology. 

\textbf{\textit{Mathematical guarantees:}} Convergence proofs for the consensus algorithm on connected or periodically-connected undirected networks, where the agents are characterized by first-order linear dynamics, are presented in \cite{Ref:Jadbabaie03,Ref:Saber2003consensus,Ref:Fax04,Ref:Tsitsiklis05,Ref:Lafferriere05,Ref:Saber07}. 
Extensions of the consensus algorithm provide convergence guarantees in \emph{directed} graphs  \cite{Ref:Ren05TAC,Ref:Ren07,Ref:Ren2007timevarying,Ref:Cao2008reaching,Ref:Qin15}, random networks \cite{Ref:Hatano2005agreement,Ref:Tahbaz2008necessary}, and time-varying networks with communication delays \cite{Ref:Murray05b,Ref:Lin09} with sampled and quantized information \cite{Ref:Olshevsky09,Ref:Nedic2010constrained,Ref:Yu13}.
Further extensions with second-order linear dynamics \cite{Ref:Ren2007IET,Ref:Ren2008double,Ref:Cao10,Ref:Kurths10},  heterogeneous, nonlinear, time-varying Lagrangian dynamics \cite{Ref:Slotine04,Ref:Slotine05c,Ref:Slotine07,Ref:Chung09_TRO}, Lipschitz nonlinearities \cite{Ref:Rezaee17}, and event-triggered control \cite{Ref:Dimarogonas2012distributed} have been studied.  
In \cite{Ref:Kingston10}, the authors prove convergence for \emph{continuous} distributions of agents. Consensus algorithms that maintain the connectivity of the communication network are proposed in \cite{Ref:Egerstedt07}. In \cite{Ref:Ji2008containment,Ref:Rahmani09,Ref:Egerstedt12,Ref:Liu2011controllability}, the authors study the stability and controllability of the consensus protocol with multiple leaders. 
Byzantine consensus algorithms for networks with Byzantine faults \cite{Ref:Lamport82,Ref:Fischer1985impossibility} with two communication steps have been developed \cite{Ref:Martin06fast}. 

For static communication graphs, the convergence rate of the consensus algorithm is proportional to the the second smallest eigenvalue of the  graph Laplacian.
The convergence speed of a wide class of linear consensus algorithms employing the consensus protocol admits a quadratic upper bound as a function of the number of agents \cite{Ref:Tsitsiklis09}, even with a time-invariant communication network. However, nonlinear consensus-based control laws can achieve consensus in finite time \cite{Ref:Cortes06b,Ref:Khoo09,Ref:Wang10}. 

\textbf{\textit{Communication bandwidth:}} At each time instant, each agent shares its own state with all its neighbors (i.e., all agents within communication reach for the prototypical consensus problem, and one agent for gossip and cyclic pursuit algorithms); thus, the resulting bandwidth use scales with the number of links in the communication graph. 

\textbf{\textit{Applications:}} \textit{Spatially-organizing behaviors}: 
Synchronization  \cite{Ref:Dorfler14,Ref:Dorfler13,Ref:Yu09pinning,Ref:Li2006global}, 
Flocking \cite{Ref:Tanner07,Ref:Tanner2003stable,Ref:Tanner2003stable2,Ref:Spong08b,Ref:Dorigo12} with collision avoidance \cite{Ref:Saber06,Ref:Cucker08}, 
Formation flying \cite{Ref:Lawton03decentralized,Ref:Ren2004JGCD,Ref:Leonard08,Ref:Chung12,Ref:Chung09,Ref:Chung09_TRO,Ref:Hadaegh13}, 
Coverage control \cite{Ref:Schwager08},
\textit{Collective decision making}: 
Localization \cite{Ref:Franceschelli14}, 
Distributed error detection and recovery \cite{Ref:Franceschelli08}. 
We refer the reader to \cite{Ref:Ren2005survey,Ref:Ren2008book,Ref:Garin10,Ref:Mesbahi2010book} for in-depth surveys of applications of the consensus algorithm.

Cyclic pursuit algorithms can be used for formation flying \cite{Ref:Marshall04,Ref:Ramirez10,Ref:Singh15} and rendezvous \cite{Ref:Ramirez10}.

Consensus-based algorithms can also be used for \textit{distributed estimation} and paired with a local controller. Consensus-based distributed estimation algorithms can be broadly classified into three categories based
on their representation of the states of the target dynamics:
\begin{itemize}
\item Algorithms that estimate only the mean and the covariance matrix of the target's states \cite{Ref:Speyer79,Ref:Borkar82,Ref:Chen02,Ref:Boyd04,Ref:Olfati2005consensus,Ref:Olfati2007Kalman,Ref:Xiao2007distributed,Ref:Hespanha2007survey,Ref:Tomlin08,Ref:Saber09,Ref:Battistelli15,Ref:Rabbat04,Ref:Murray05,Ref:Smith2007,Ref:Freeman08}.
\item Algorithms that reach an agreement across the sensor network over a discrete set of hypotheses about the states of the target \cite{Ref:Pavlin10,Ref:Jadbabaie12,Ref:Nedic14_Cesar}.
\item Algorithms that estimate the posterior probability distribution of the states of the target \cite{Ref:Bailey12,Ref:Ahmed13,Ref:Fraser12,Ref:Hlinka12,Ref:Hlinka14,Ref:Battistelli14,Ref:Bandyopadhyay14_ACC_BCF,Ref:Bandyopadhyay18_Auto}.
\end{itemize}
Algorithms in the last class can be used for estimation over continuous domains and incorporate nonlinear target dynamics, heterogeneous nonlinear measurement models, and non-Gaussian uncertainties.

\section{Artificial Potential Functions (APF)}
\label{sec:potential-functions}

\begin{figure}
\centering
\begin{subfigure}[h]{.5\textwidth}
\includegraphics[width=\textwidth]{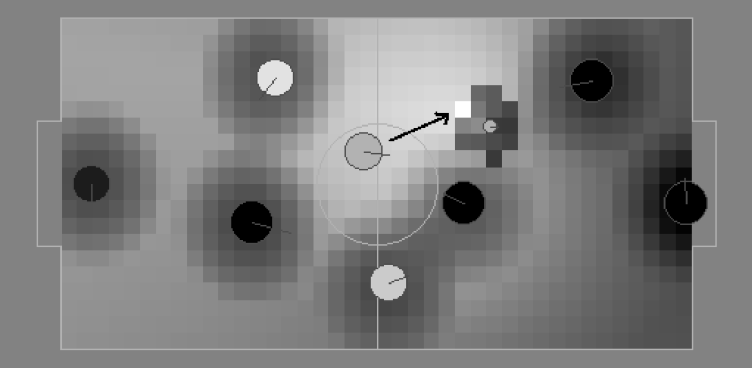}
\end{subfigure}

\vspace{.2em}

\begin{subfigure}[h]{.5\textwidth}
\includegraphics[width=\textwidth]{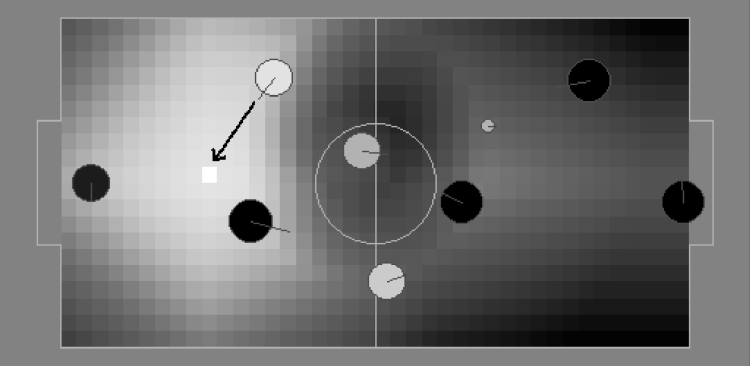}
\end{subfigure}

\vspace{.2em}

\begin{subfigure}[h]{.5\textwidth}
\includegraphics[width=\textwidth]{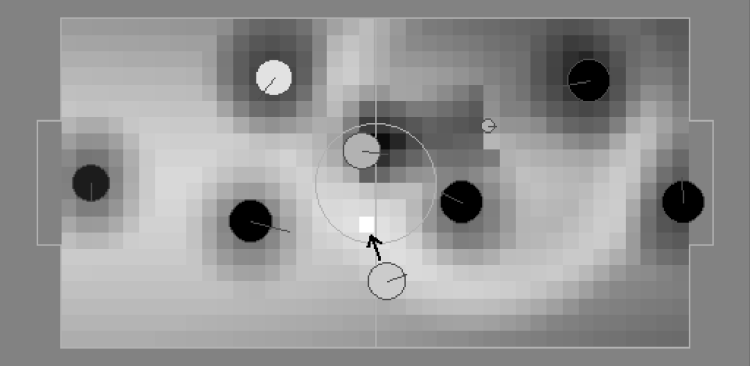}
\end{subfigure}
\caption{An example of \textbf{artificial potential functions} (APFs): potential fields for determining the active position (top), strategic position (center), and support position (bottom) of robotic RoboCup players. APF algorithms synthesize agents' control inputs using the gradient of a suitably-defined potential function, where goals act as attractors, and obstacles have a repulsive effect. \copyright 2002 IEEE. Reprinted, with permission, from \cite{Ref:Weigel02}.}
\label{fig:apf:freiburg}
\end{figure}

APF algorithms are a family of control techniques where the agents' control inputs  are synthesized on the basis of the gradient of a suitably-defined potential function \cite{Ref:Khatib86}. 
Potential functions  where the goal position (or positions) is guaranteed to be reachable from any feasible initial state are known as \emph{navigation functions} \cite{Ref:Koditschek90,Ref:Rimon92}.
\revXXIa{
Figure \ref{fig:apf:freiburg} (from \cite{Ref:Weigel02}) shows a prototypical set of artificial potential functions used for coordination of robotic soccer players in the RoboCup.
}

\textbf{\textit{Mathematical description:}}
In the prototypical APF algorithm, each agent updates its position using the following equation:
\begin{equation}
x_{k+1}^{i} = x_{k}^{i} - \nabla_{x_{k}^{i}} J\thinspace, \qquad \forall i\in\mathcal{V}\thinspace, \thinspace\forall k\in\mathbb{N}\thinspace, 
\end{equation}
where
\begin{equation*}
J = f_{\text{attraction}}(x_{k}^{i}, \text{target}) + f_{\text{repulsion}}(x_{k}^{j} \thinspace \forall j\in\mathcal{V}, \text{obstacles}) \thinspace. 
\end{equation*}
The nonlinear function $f_{\text{attraction}}$ creates a potential field that attracts the agent to its goals, and the nonlinear function $f_{\text{repulsion}}$ creates a potential field that repels the agent from other agents and obstacles in the environment \cite{Ref:Merheb16}. 

\textbf{\textit{Mathematical guarantees:}}
Convergence of navigation function algorithms can be proved using LaSalle's Invariance Principle \cite{Ref:Dimarogonas05APF,Ref:Dimarogonas06}. 
Conversely, the more general class of APF algorithms do not generally offer strong analytical guarantees due to the possible presence of local minima. We refer the reader to \cite{Ref:Koren91} for a thorough discussion of the limitations of APF.

\textbf{\textit{Communication bandwidth:}} 
Distributed implementations of APF techniques generally offer excellent scalability, requiring agents to communicate with a limited number of their immediate neighbors \cite{Ref:Vadakkepat01}.

\textbf{\textit{Applications:}}
\textit{Spatially-organizing behaviors}: Formation control and pattern formation \cite{Ref:Leonard2001virtual,Ref:Egerstedt2001formation,Ref:Tanner05,Ref:Gazi05,Ref:De2006formation,Ref:Leonard07,Ref:Leonard07b,Ref:Hsieh08,Ref:Spong08,Ref:Barnes09swarm,Ref:Tanner12} in satellites \cite{Ref:Pinciroli08,Ref:Saaj06spacecraft,Ref:Peck07spacecraft,Ref:Bandyopadhyay12_SMC}, 
Shape formation \cite{Ref:Slotine09},
Flocking \cite{Ref:Chuang07,Ref:Zavlanos2007flocking},
Perimeter formation \cite{Ref:Bruemmer02},
Coverage control  \cite{Ref:Hussein07,Ref:williams13},
Sensor network optimization \cite{Ref:Howard2002mobile,Ref:Ogren2004cooperative}.
\textit{Collective explorations}: Area exploration \cite{Ref:Xu11}, 
Foraging \cite{Ref:Gazi2003stability,Ref:Passino04,Ref:Gazi07}. 
\textit{Collective decision making}: Multi-agent path planning \cite{Ref:Koditschek90,Ref:Loizou02,Ref:Loizou06,Ref:Dimarogonas03,Ref:Dimarogonas05,Ref:Lionis07,Ref:Warren90,Ref:Lee13,Ref:Vadakkepat01,Ref:Ayanian10},
Multi-agent cooperative manipulation \cite{Ref:Khatib96,Ref:Bai10},
Network connectivity maintenance \cite{Ref:Dani12,Ref:Dimarogonas08,Ref:Zavlanos07,Ref:De2006decentralized},
Task allocation \cite{Ref:Weigel02,Ref:Pappas08b}.

\section{Distributed Feedback Control}
\label{sec:distributed-feedback}
\begin{figure}
\centering
\begin{subfigure}[h]{\ifonecolumn .48 \else .35\fi \textwidth}
\includegraphics[width=\textwidth]{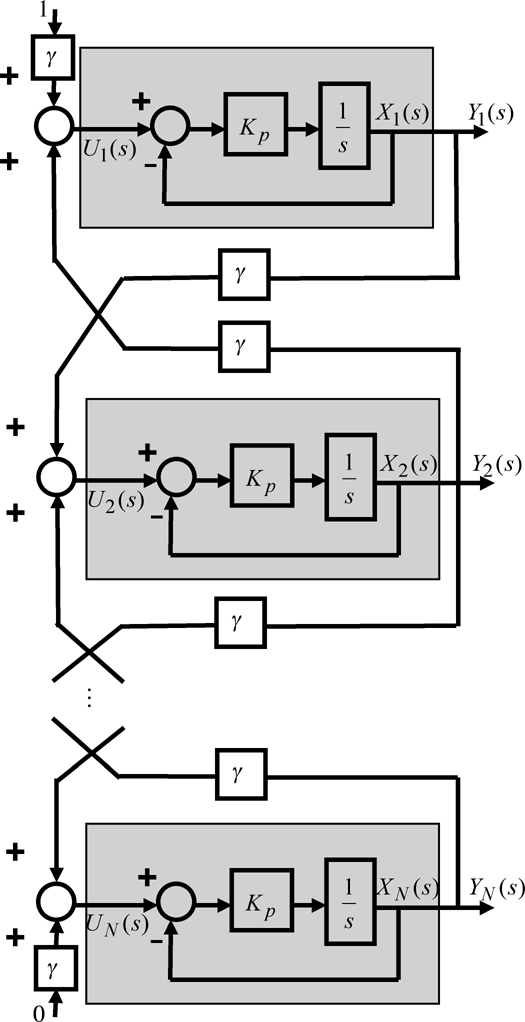}
\end{subfigure}
\begin{subfigure}[h]{\ifonecolumn .48 \else .4 \fi \textwidth}
\includegraphics[width=\textwidth]{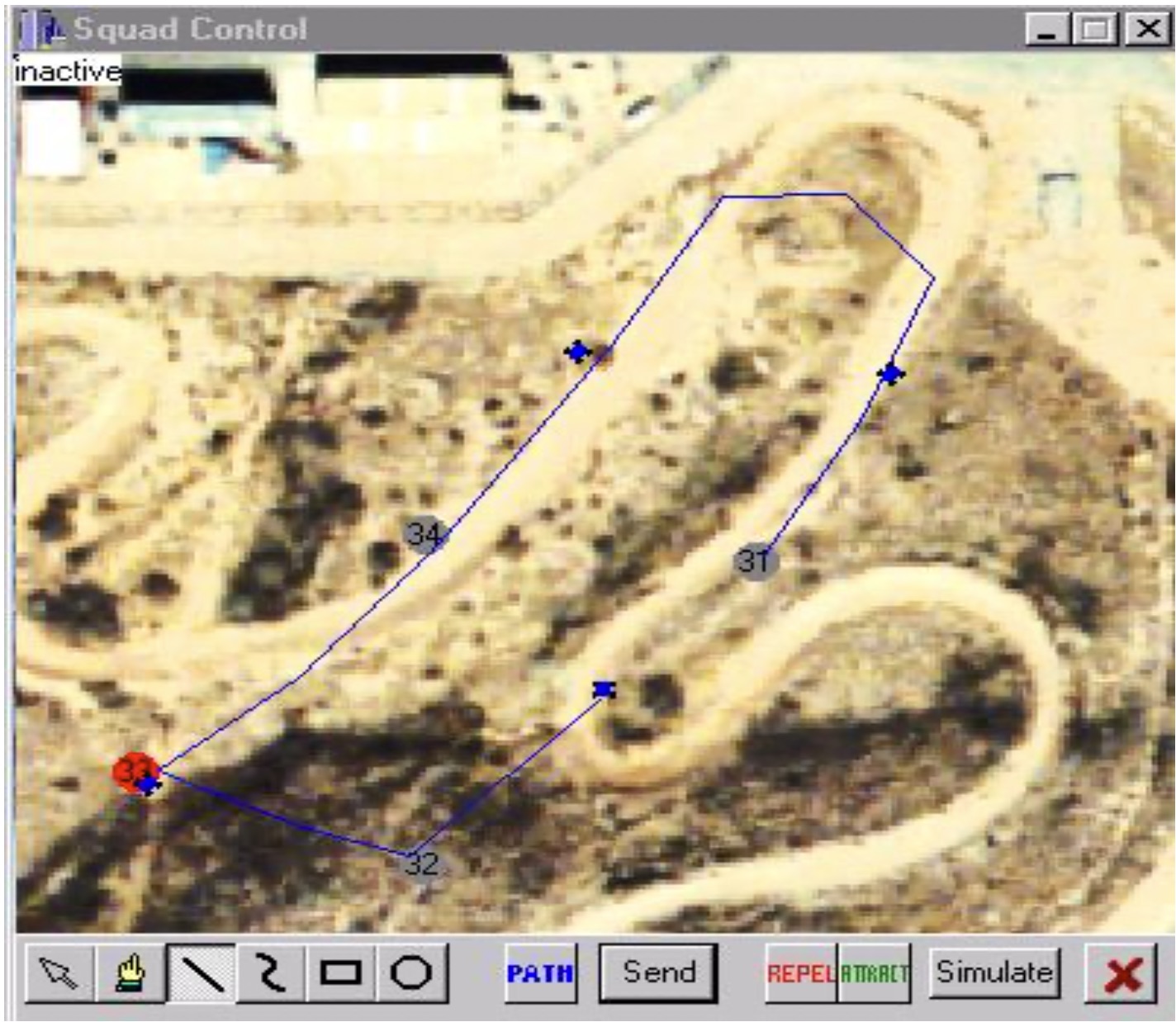}
\end{subfigure}

\caption{Example of \textbf{distributed feedback control}. 
\ifonecolumn Left\else Top\fi: Control block diagram for a $N$-vehicle interaction problem. \ifonecolumn Right\else Bottom\fi: perimeter being guarded by robot sentries. The distributed feedback controller ensures that the vehicles space themselves along the perimeter, while avoiding collisions. 
In distributed feedback control algorithms, each agent is endowed with a classical feedback controller whose input is the concatenation of the agent's output and the outputs of its neighbors. The overall system behaves as a dynamical feedback system, where the sparsity pattern of the controller reflects the
topology of the agent's communications.
\copyright 2002 IEEE. Reprinted, with permission, from \cite{Ref:Feddema02}.}
\label{fig:feedback:patrolling}
\end{figure}

In distributed feedback control algorithms (often referred to as \emph{decentralized} feedback control algorithms in the literature), each agent is endowed with a classical feedback controller whose input is the concatenation of the agent's output and the outputs of its neighbors. The overall network can be modeled as dynamical feedback system where the sparsity pattern of the controller reflects the topology of the agent's communication network.
\revXXIa{
Figure \ref{fig:feedback:patrolling} (from \cite{Ref:Feddema02}) shows a prototypical control block diagram for a multi-vehicle system and an application of distributed feedback control to multi-robot patrolling.
}

\textbf{\textit{Mathematical description:}}
Each agent is endowed with a feedback controller that uses the agent's and its neighbors' outputs as the input
 \cite{Ref:Bamieh02,Ref:Feddema02}.
Formally, every agent synthesizes its control action $u^i$ as:
\begin{equation}
u^i_{k+1}=f^i(y^i_k, \cup_{j\in \mathcal{N}_{k}^{i}} y^j_k)
\end{equation}
where $f^i(y^i_k, \cup_{j\in \mathcal{N}_{k}^{i}} y^j_k)$ is
\revXXIa{
 a linear dynamical system, or the concatenation of an observer and a linear controller.
 }
For instance, in the case of static feedback control, $f^i(y^i_k, \cup_{j\in \mathcal{N}_{k}^{i}} y^j_k)=K^{ii}y^i_k + \sum_{j\in \mathcal{N}_{k}^{i}} K^{ij}y^j_k$ is a linear function of the agents' observations; in the case of a distributed LQG controller, the function represents the concatenation of a Kalman filter and a LQR controller.

\textbf{\textit{ Mathematical guarantees:}}
Distributed feedback control has seen an incredible amount of interest in the controls community, since the structure of the resulting multi-agent system (considered as a whole) can readily be analyzed with well-understood control-theoretical tools. We refer the reader to \cite{Ref:Sandell1978survey} and \cite{SiljakZecevic2005}  for comprehensive surveys and to \cite{Ref:Feddema02} for a robotic application.

In particular, a number of techniques have been developed to synthesize distributed optimal controllers that minimize the $\mathcal{L}_2$ norm of a desired objective (LQG controllers).
The works in \cite{Ref:DAndrea03,Ref:DAndrea03b,Ref:Langbort2004distributed,QiSalapakaEtAl2004,Rantzer2006,Ref:Rotkowitz06,Rantzer2006b,Ref:Swigart10,SwigartLall2010b,SwigartLall2010,LamperskiDoyle2011,LamperskiDoyle2012} propose tools based on linear matrix inequalities (LMI) to solve the distributed LQG synthesis problem as a convex program for linear and nonlinear systems in presence of communication delays and uncertainty.

Tools for synthesis of distributed LQG control are available that can adapt to noisy communication links;  the work in \cite{Ref:Sahai06} studies the fundamental limitations of performance of the noisy LQG problem and identifies capacity limits of noisy channels that are required to ensure stabilizability.

In \cite{Ref:Borrelli08}, the authors propose an analytical solution to the LQR problem in the case where identical subsystems are interconnected. The paper \cite{Ref:Zhang12} proposes LQR and adaptive control techniques on \emph{directed} graphs.

A closely related area of research studies the design of controllers for \emph{infinite-dimensional} systems with distributed parameters (e.g., \cite{Ref:Bamieh02,Ref:Motee2008optimal}), with applications including the control of very large-scale multi-agent systems such as vehicular platoons.

\textbf{\textit{Communication bandwidth:}}
At every time step, agents communicate their state to their neighbors; thus, the bandwidth use of the algorithm is proportional to the number of links in the communication graph.%

 \textbf{\textit{Applications:}}
Distributed control has been widely used for a variety of applications, including distributed estimation and tracking \cite{Ref:Rotkowitz06,Ref:Drew05,Ref:Sahai06,Ref:Garone11,Ref:Liu04,Ref:Zhang12,Ref:Borrelli08}, 
stabilization of distributed systems \cite{Ref:Wang73}, 
and formation flying \cite{Ref:Ogren02,Ref:Stipanovic04decentralized}.
Distributed control of infinite-dimensional spatially distributed systems has been applied to multi-agent trajectory optimization \cite{Ref:Ferrari14}, multi-agent formation \cite{Ref:Krstic15}, particle swarm control \cite{Ref:Allain14optimal}, and stability analysis with integro-differential equation \cite{Ref:Mogilner99}.

\section{Geometric Algorithms}
\label{sec:geometric-algo} 
In geometric algorithms, agents use their neighbors' location and velocity information to perform spatially organizing tasks and path planning. Communication with neighbors can often be replaced by sensory information (e.g. from camera sensors) resulting in highly scalable algorithms.

\subsection{Voronoi-based Algorithms \label{subsec:Voronoi}}
\begin{figure}
\centering
\begin{subfigure}[h]{\ifonecolumn .48\else .4\fi\textwidth}
\includegraphics[width=\textwidth]{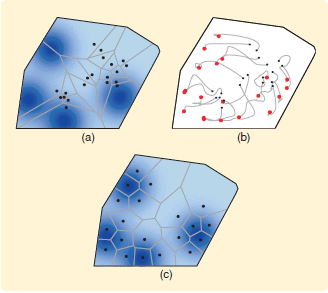}
\end{subfigure}
\begin{subfigure}[h]{\ifonecolumn .48\else .4\fi\textwidth}
\includegraphics[width=\textwidth]{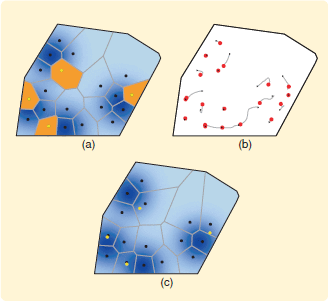}
\end{subfigure}
\caption{
An example of \textbf{geometric algorithms}.
Left: Voronoi algorithm for coverage control. Each of the 20 mobile agents moves toward the centroid of its Voronoi cell. Areas of the convex polygon with greater importance are colored in darker blue.
Right: Adaptive network behavior under agent failures. After the final configuration in the figure on the left is reached, four network agents (yellow) fail and cease to provide coverage in their respective Voronoi cells (orange). The rest of the network adapts to the new situation. (a) and (c) show the initial and final agent locations with the corresponding Voronoi partitions. (b) shows the gradient descent flow.
In geometric algorithms, agents use their neighbors' location and speed information (often obtained through sensory information, with minimal communication) to make decisions and perform spatially organizing tasks and path planning.
\copyright 2007 IEEE. Reprinted, with permission, from \cite{Ref:Bullo08}.}
\label{fig:geometric:voronoi}
\end{figure}

A Voronoi diagram partitions the space into regions based on the distance from the individual agents.
\revXXIa{
The diagram can be efficiently computed in a distributed manner and used to achieve optimal coverage of a bounded region.
Figure \ref{fig:geometric:voronoi} (from \cite{Ref:Bullo08}) provides a pictorial representation of a Voronoi algorithm for coverage control, and shows how the algorithm can autonomously replan the robots' positions in response to agent failures.
}

\textbf{\textit{Mathematical description and guarantees:}} 
In coverage algorithms based on Voronoi-partitions \cite{Ref:Bullo04,Ref:Bullo08,Ref:Bullo09book}, each agent first computes its Voronoi partitions using:
\begin{equation}
\mathcal{V}_k^i = \{ q \in \mathcal{X} \mid \| q - x_k^i \|_2 \leq \| q - x_k^j \|_2 \thinspace, \forall j \neq i \} \thinspace ,
\end{equation}
where $\mathcal{X} \subset \mathbb{R}^{n_{x}}$ represent the space to be covered by these agents.
Then each agent moves towards the centroid $\mathcal{C}_{\mathcal{V}_k^i} = \frac{\int_{\mathcal{V}_k^i} q \thinspace dq}{\int_{\mathcal{V}_k^i} dq}$ of its Voronoi.
The agents asymptotically achieve an optimal coverage of the space because the centroidal-Voronoi configurations are fixed points of this algorithm \cite{Ref:Bullo04}.

\textbf{\textit{Communication bandwidth:}} 
These algorithms offer excellent scalability, since the agents only communicate with their immediate neighbors.

\textbf{\textit{Applications:}} 
\textit{Spatially-organizing behaviors}: coverage \cite{Ref:Bullo04,Ref:Cortes2005spatially,Ref:Cortes2005coordination,Ref:Bullo08,Ref:Bullo09book,Ref:Schwager07,Ref:Bhattacharya14,Ref:Martinez07a,Ref:Martinez07b,Ref:Gao08,Ref:Pappas13},
flocking \cite{Ref:Johansson05}.
\textit{Collective explorations}: area exploration \cite{Ref:Guruprasad11}.
\textit{Collective decision making}: multi-agent path planning \cite{Ref:Beard02,Ref:Sud08,Ref:Bandyopadhyay14MSC,Ref:Zhou17},
task allocation \cite{Ref:Pavone11},
distributed estimation \cite{Ref:Martinez2006optimal}.

\subsection{Circumcenter Algorithm \label{subsec:circumcenter}}
In the circumcenter algorithm, each agent moves towards the circumcenter of the point set comprised of its inclusive neighbors \cite{AndoOasaEtAl1999,Ref:Cortes06,Ref:Dimarogonas07,LinMorseEtAl2007} to rendezvous with other agents. 

\subsection{Bearing-only Algorithms \label{subsec:bearing-only}}
In bearing-only algorithms, each agent maintains its neighbor at a desired bearing angle in order to achieve a formation \cite{Ref:Fredslund02}, rendezvous \cite{Ref:Yu08rendezvous}, or cooperative localization \cite{Ref:Sharma12}.

\subsection{Maze Searching Algorithms\label{subsec:maze-searching}}
Maze-searching algorithms perform distributed multi-agent path planning through unknown stationary obstacles with no inter-agent communication \cite{Ref:Lumelsky97}, extending single-agent maze searching techniques \cite{Ref:Lumelsky87}.

\subsection{Leader-Follower Algorithms\label{subsec:leader-follower}}
In leader-follower algorithms, agents are logically arranged in a tree structure; each agent's control only depends on its own state and on the parent agent's state.
Leader-follower algorithms have been used for formation flying \cite{Ref:Mesbahi99formation,Ref:Mesbahi01,Ref:Egerstedt2001formation,Ref:Beard00feedback,Ref:Consolini09} and formation control \cite{Ref:Desai1998controlling,Ref:Desai01modeling,Ref:Das2002vision}; the work in \cite{Ref:Cheng10} proposes a robust leader-follower algorithm based on neural networks, and 
\cite{Ref:Tanner2004leader} studies the stability of leader-to-formation schemes.

\subsection{Velocity Obstacle (VO) based Algorithms \label{subsec:velocity-obstacle}}
Velocity obstacles-based algorithms command an agent's velocity so as to avoid collisions with other mobile agents based on (i) observations of the other agents' locations and velocity and (ii) deterministic assumptions about their behavior. Different VO algorithms differ in the assumptions regarding the behavior of other agents: classical VO algorithms (also known as maneuvering-board algorithms) assume that other agents will maintain a constant velocity, whereas reciprocal velocity obstacle (RVO)-based algorithms assume that all agents will perform evasive maneuvers according to the same RVO algorithm. 

VO and RVO collision avoidance algorithms are widely used in  multi-agent motion planning \cite{Ref:vabdenBerg08,Ref:Van2011reciprocal,Ref:Bareiss15,Ref:Hoy14}.

\section{State Machines and Behavior Composition}
\label{sec:automata-algo}

The algorithms presented in the previous sections generally rely on \emph{continuous} controllers to steer the behavior of the system. In a departure from this approach, state machines and behavior composition algorithms rely on \emph{discrete} descriptions of the agent's states and actions and design complex orchestrations of transitions to achieve a desired global behavior. For this reason, state machines and behavior composition algorithms are especially popular for combinatorial cooperative decision-making problems such as task allocation. %

\begin{figure}
\centering
\begin{subfigure}[h]{.48\textwidth}
\includegraphics[width=\textwidth]{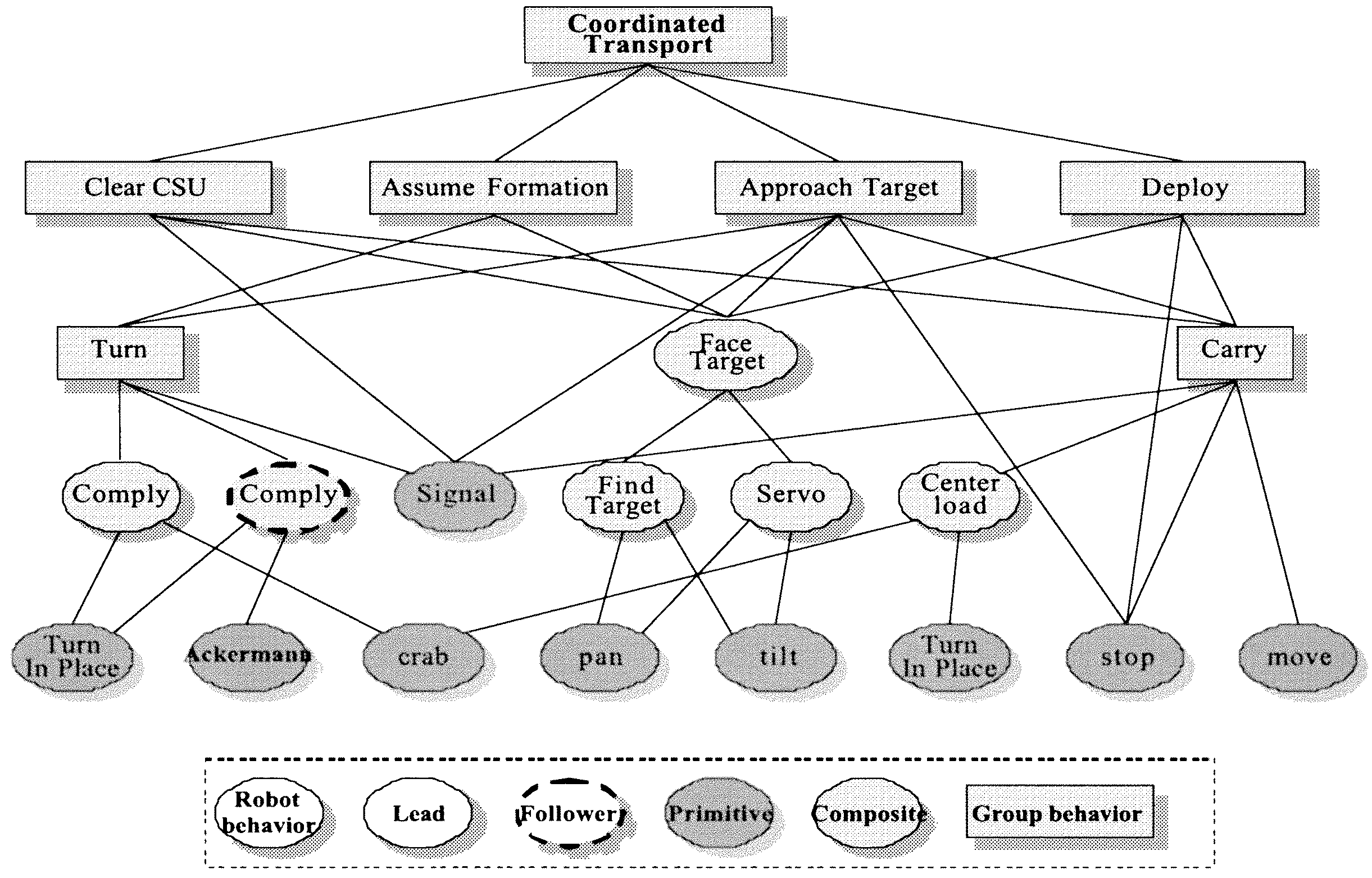}
\end{subfigure}
\begin{subfigure}[h]{.48\textwidth}
\includegraphics[width=\textwidth]{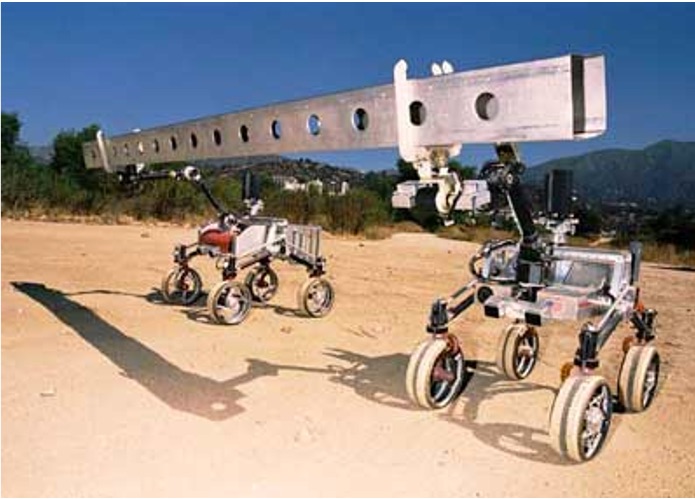}
\end{subfigure}
\caption{
A \textbf{behavior composition} algorithm in action.
Left: Behavior hierarchy describing a coordinated transport task. Bubbles represent single robot behaviors and boxes represent group behaviors. The hierarchy shows how the behaviors are composed from lower-level behaviors.
Right: Technology concept robotic work crew using the behavior composition algorithm to cooperatively transport a beam.
State machines and behavior composition algorithms rely on discrete descriptions of the agent's states and actions, and design complex orchestrations of transitions and message exchanges to achieve a desired global behavior.
\copyright 2003 IEEE. Reprinted, with permission, from \cite{Ref:Huntsberger03}.}
\label{fig:behavior:campout}
\end{figure}

\revXXIa{
Figure \ref{fig:behavior:campout} (from \cite{Ref:Huntsberger03}) shows a behavior composition architecture designed for cooperative transport, and a hardware demonstration of the architecture in support of planetary surface exploration.
}

\subsection{Automata-based Algorithms for Agreement and Leader Election\label{subsec:message-passing}}
\textbf{\textit{Mathematical description:}}
In automata-based distributed algorithms, each agent is modeled as an input/output automaton characterized by four properties \cite{Ref:Burns80}, \cite[Ch. 2]{Ref:Lynch97}:
\begin{itemize}
\item A (possibly infinite) set of \emph{states}.
\item A set of \emph{starting states};
\item A \emph{message-generating function} that determines the messages that the agent will send to its neighbors based on the agent's state and  neighbors.
\item A \emph{state-transition function} that maps the agent's state and the messages it receives to a new state.
\end{itemize}
Agents exchange messages across communication links, which may drop messages. Each agent can also experience stopping failures as well as \emph{Byzantine} failures (where it can perform arbitrary state transitions and generate arbitrary messages).

We specifically focus our attention on automata-based algorithms for consensus and agreement problems \cite{LamportShostakEtAl1982}, leader election \cite{FredericksonLynch1987} and spanning tree construction \cite{Ref:Gallager83,Ref:Awerbuch87}. We refer the interested reader to \cite{Ref:Lynch97} for a thorough survey.

\textbf{\textit{Mathematical guarantees:}}
In general, automata-based algorithms offer strong correctness guarantees in presence of static communication networks and in absence of link failures. Even very simple coordination problems cannot be solved by \emph{deterministic} automata-based algorithms if communication links can experience arbitrary failures \cite{Gray1978}; however, randomized algorithms are available that can overcome this difficulty with high probability \cite{VargheseLynch1992}.
Critically, automata-based algorithms are also available that can solve agreement problems in presence of limited \emph{Byzantine} failures as well as stopping failures \cite{PeaseShostakEtAl1980, LamportShostakEtAl1982, Dolev1982} -- a capability seldom found in other classes of  mathematical techniques.

A number of algorithms are also available to cope with \emph{asynchronous} multi-agent systems where the agents do not share a common clock. Such algorithms are beyond the scope of this review: we refer the reader to \cite[Part II]{Ref:Lynch97} for a comprehensive review.

\textbf{\textit{Communication bandwidth:}} 
Different automata-based algorithms can exhibit vastly different bandwidth usage, ranging from $O(1)$ to $O(|\mathcal{V}||E|)$. However, in most cases of practical interest, automata-based algorithms can be tuned to use very low amounts of bandwidth (and, more generally, very low numbers of messages) at the price of additional time complexity (see, for instance, the discussion in \cite{FredericksonLynch1987}). 

\textbf{\textit{Applications:}}
\cite{RossiPavone2013,Ref:Rossi14} propose applications to multi-agent systems with communication bandwidth constraints. Applications of such algorithms include distributed estimation and task allocation; in addition, automata-based algorithms can be used as a building block to establish communication infrastructures that can be leveraged by other coordination algorithms (e.g., an automata-based leader-election algorithm can be used to identify a leader and establish lightweight communication infrastructure in a multi-agent system, enabling the distributed implementation of centralized coordination algorithms). Simple automata-based algorithms can be used to estimate the locations of a network of agents from trilateration, a fundamental building block for distributed estimation applications \cite{Ref:Nagpal2003organizing}.

\subsection{Behavior Composition\label{subsec:behavior-composition}}
Behavior composition algorithms rely on composition of elementary behaviors (e.g. "push in a given direction" or "grasp") through prescribed switching rules; applications include collective transport and task allocation
\ifjournalv\cite{MataricNilssonEtAl1995,Ref:Rus1995moving,Ref:Parker98,Ref:Werger00broadcast,Ref:Huntsberger03, Ref:Wolf17}\else\cite{Ref:Rus1995moving}\fi.

\subsection{Petri Networks\label{subsec:petri-nets}}
Petri networks and workflow network are a popular modeling technique to describe the set of possible states and actions of a multi-agent systems, and enable the efficient computation of system properties such as liveness and reachability of certain states \cite{Ref:King03,Ref:Kotb12}. In multi-agent systems, they are used for centralized task allocation  and recovery from deadlock/livelock.

\subsection{Resource Allocation Systems\label{subsec:RAS}}
Resource Allocation System algorithms can be used to generate collision-free multi-agent motion plans. The agents' task space is partitioned in a number of non-overlapping regions, treated as \emph{resources}; the resources are then allocated to the agents to ensure that no two agents can occupy the same region simultaneously \cite{Ref:Reveliotis11}.

\section{Bio-Inspired Algorithms}
 \label{sec:bio-inspired-algo}

Bio-inspired algorithms mimic the behavior of swarms of animals such as insects and fish. 
They are often similar in spirit to state machines and behavior composition algorithms: their defining characteristics, in addition to their biology-inspired heritage, include the use of very limited communication and on-board computation resources \revXXIa{and the emphasis on simple local rules and algorithms designed to elicit an emergent group behavior --- characteristics} that typically result in low bandwidth use and excellent scalability. 

\revXXIa{
A very large number of bio-inspired algorithms have been proposed in the literature.
In this section, we present a small and opinionated selection, with the goals of providing a representative sample of the literature and of identifying the fundamental characteristics shared across algorithms in this class.
}

We remark that this paper focuses on bio-inspired algorithms with direct multi-agent robotic applications: accordingly,  bio-inspired \emph{global optimization} algorithms such as genetic algorithm \cite{Ref:Horn1994niched,Ref:Deb2002fast},  particle swarm optimization \cite{Ref:Kennedy2011particle,Ref:Poli2007particle} and ant colony optimization \cite{Ref:Dorigo96ant,Ref:Dorigo2006Ant,Ref:Dorigo97ant} are beyond the scope of this paper.

\subsection{Stigmergy Collective Construction Algorithm\label{subsec:stigmergy}} 
In termite-inspired collective construction algorithms, agents coordinate their activity by observing and modifying the shared environment (e.g., adding or moving material) according to simple rules, with no explicit communication - a process known as \emph{stigmergy} in the biology literature.
Synthesis algorithms are available to design sets of simple rules that result in collective construction of a desired structure \cite{Ref:Nagpal14termite,Ref:Nagpal06,Ref:Terada04stigmergy}.

\begin{figure}
\centering
\includegraphics[width=\columnwidth]{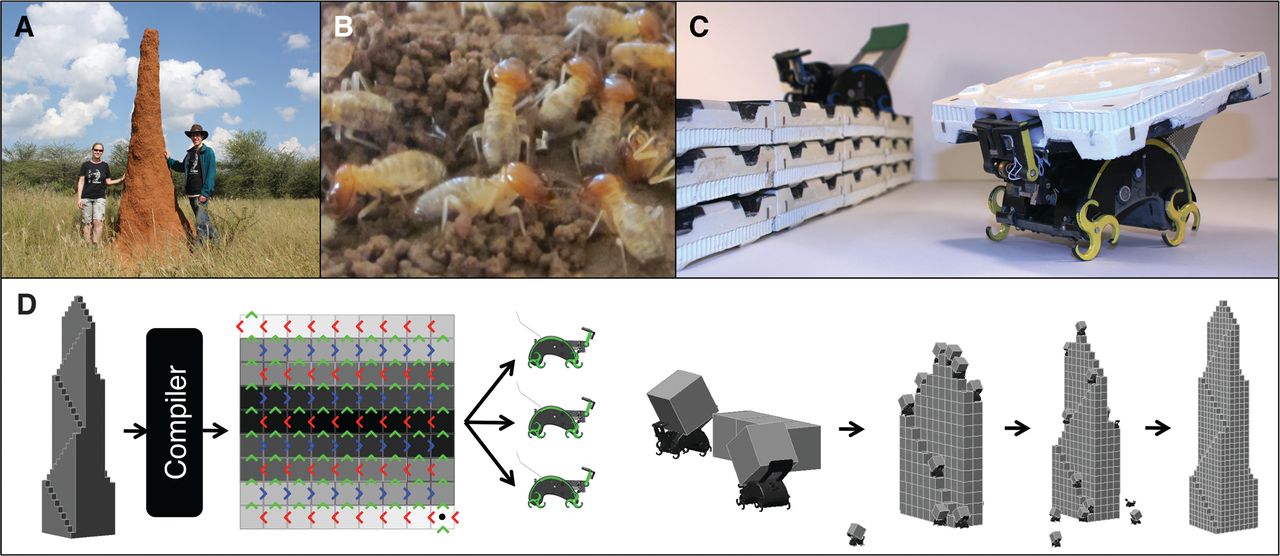}
\caption{
\textbf{Bio-inspired} stigmergy collective construction algorithm in action. Complex meter-scale termite mounds (A) are built by millimeter-scale insects (B), which act independently with local sensing and limited information. (C) shows a physical implementation of the proposed system, with independent climbing robots that build using specialized bricks. (D) shows a system overview for building a specific predetermined result: A user specifies a desired final structure; an offline compiler converts it to a "structpath" representation, which is provided to all robots; robots follow local rules that guarantee correct completion of the target structure.
Bio-inspired algorithms mimic the behavior of swarms of animals such as insects and fish. They are often similar in spirit to state machines and behavior composition algorithms: their defining characteristics are (i) their biology-inspired heritage and (ii) the use of very limited communication and on-board computation resources, typically resulting in low bandwidth use and good scalability.
From \cite{Ref:Nagpal14termite}. Reprinted with permission from AAAS.
}
\label{fig:bioinspired:stigmergy}
\end{figure}

\revXXIa{
Figure \ref{fig:bioinspired:stigmergy} (from \cite{Ref:Nagpal14termite}) shows the inspiration and architecture of the stigmergy algorithm, and a hardware demonstration with individual robots carrying construction material.
}

\subsection{Kilobot Self-Assembly Algorithm \label{subsec:kilobot}}
The Kilobot self-assembly algorithm \cite{Ref:Nagpal14} enables a swarm of a thousand robots to self-organize in complex two-dimensional shapes using inter-agent communication for localization, edge-following for path planning, and a gradient-based formation control routine inspired by the behavior of simple multi-cellular organisms.

\subsection{Optimotaxis and Immune-Inspired Source-Searching Algorithm \label{subsec:optimotaxis}}

The optimotaxis algorithm, inspired by the run and tumble behaviors of bacteria (chemotaxis), enables a swarm of agents with no communication or localization capabilities to locate the source of a signal of interest \cite{Ref:Hespanha08}. Agents move  according to a random walk induced by the concentration of the signal; the final distribution of the agents coincides with the location of the signal source.
\revXXIa{
The behavior can be improved by using a L\'{e}vy walk \cite{Nurzaman2008Yuragi}, which approximates the search behavior of T-cells in the immune system \cite{fricke2016immune}.
}

\subsection{Beeclust Foraging Algorithm \label{subsec:beeclust}}
The Beeclust foraging algorithm, inspired by the behavior of honey bees, searches for signal peaks in the environment by making agents wait at each given location for a time proportional to the signal concentration at that location \cite{Ref:Hereford10,Ref:Hereford11}. 

\subsection{Shepherding Algorithm\label{subsec:shepherding}}

The shepherding algorithm in  \cite{Ref:Strobom14} uses simple behavioral primitives inspired by the behavior of dogs herding flocks of sheep. These behavior primitives enable a small number of controlled agents to control  the spatial distribution of large numbers of uncontrolled agents.

\subsection{Gillespie Self-Assembly Algorithm\label{subsec:gillespie}}
In \cite{Ref:Ayanian08stochastic}, the authors propose a stochastic, distributed algorithm for the self-assembly of a group of modular robots into a geometric shape; the algorithm is inspired by the Gillespie algorithm for chemical kinetics.

\subsection{Mergeable Modular Robots\label{subsec:MNS}} 
The works in \cite{Ref:Dorigo17,Ref:Dorigo06,Ref:Mondada2004swarm,Ref:Dorigo2013swarmanoid,Ref:Kotay1998self} propose bio-inspired algorithms that enable groups of robots to connect to form larger bodies or split into separate bodies, with independent controllers, and self-heal by removing or replacing malfunctioning body parts.

\section{Density-Based Control Algorithms}
\label{sec:density-based-control}
In the algorithms discussed in previous sections, agents are represented in a \emph{Lagrangian} framework; that is, the state of each agent is tracked individually. In contrast, the density-based control algorithms discussed in this section treat  a swarm of agents as a continuum, and control the density of agents in the workspace in an \emph{Eulerian} framework.  

\subsection{Markov Chain (MC) based Algorithm \label{subsec:markov-chain}}
Markov chain-based algorithms are highly applicable to guidance of large swarms ($10^3-10^6$ agents), where it is not practical for the agents to share their individual states.  

\begin{figure}
\centering
\includegraphics[width=\columnwidth]{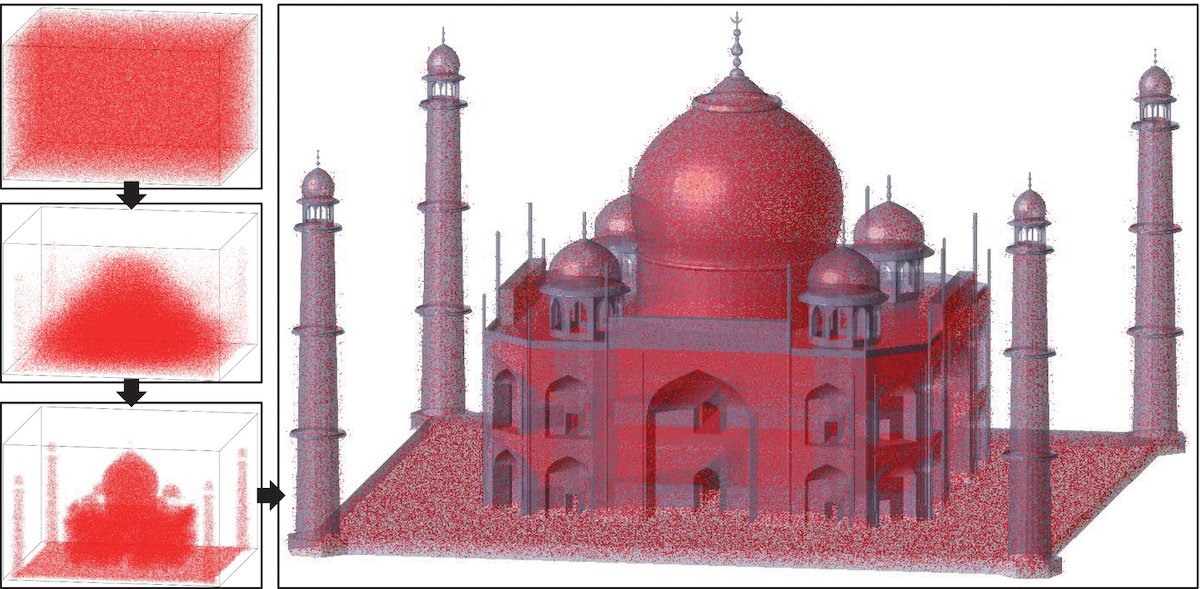}
\caption{\textbf{Density-based control} algorithms adopt an Eulerian framework by treating agents as a continuum and devising control laws to drive the agents' density towards a specified distribution.
In the figure, using an inhomogeneous Markov Chain algorithm for shape formation, a million swarm agents (shown in red) attain the complex three-dimensional (3-D) shape of the Taj Mahal (translucent silhouette shown in gray) in simulation. The physical space is partitioned into $100\times 100\times 70$ bins.
\copyright 2017 IEEE. Reprinted, with permission, from \cite{Ref:Bandyopadhyay17_TRO}.
}
\label{fig:mc:tajmahal}
\end{figure}

\revXXIa{
Figure \ref{fig:mc:tajmahal} (from \cite{Ref:Bandyopadhyay17_TRO}) shows an inhomogeneous Markov Chain algorithm steering a million-agent swarm to reproduce a complex physical shape in simulation.
}

\textbf{\textit{Mathematical description and guarantees:}} 
The workspace $\mathcal{X} \subset \mathbb{R}^{n_x}$ is partitioned into disjoint bins (cells), which are represented as $B[\ell], \forall \ell \in \{1,\ldots,n_{\text{bin}}\}$. 
The transition probability between bins for the $i^{\text{th}}$ agent at the $k^{\text{th}}$ time instant is dictated by the Markov matrix $\mathbf{M}_k^i$. For example, the $i^{\text{th}}$ agent's transition probability from bin $B[\ell]$ to bin $B[m]$ at the $k^{\text{th}}$ time instant is given by:
\begin{equation}
\mathbf{M}_k^i[\ell, m] = \mathbb{P}(i^{\text{th}} \text{ agent transitions from } B[\ell] \text{ to } B[m]) \thinspace .
\end{equation}
The entries in the Markov matrix $\mathbf{M}_k^i$ are selected to ensure that, if agents follow the transition probabilities in the matrix, the swarm  converges to the desired distribution with an exponential rate \cite{Ref:Bandyopadhyay17_TRO}. 

\textbf{\textit{Communication bandwidth:}} 
Homogeneous MC algorithms require no communication \revXXIa{once the Markov matrix is known} \cite{Ref:Acikmese12,Ref:Acikmese15asian}, while inhomogeneous MC algorithms \cite{Ref:Bandyopadhyay17_TRO} (where the transition probabilities $\mathbf{M}^i_k$ are time-varying) use the consensus algorithm to estimate the current swarm distribution. 

\textbf{\textit{Applications:}} 
Both homogeneous \cite{Ref:Milutinovic06,Ref:Acikmese12,Ref:Acikmese15asian,Ref:Chattopadhyay09,Ref:Acikmese15convex,Ref:Acikmese15_AR,ChamieAcikmese2016,DemirerElChamie2017} and inhomogeneous MC algorithms \cite{Ref:Bandyopadhyay_IROS16,Ref:Bandyopadhyay17_TRO,Ref:Giri14,Ref:Jang2018local} algorithms can be used for pattern formation, coverage, area exploration, and goal searching. 
Additional applications include multi-agent surveillance \cite{Ref:Grace05stochastic}, coverage \cite{Ref:Leonard13,Ref:Gundry12markov}, and task allocation \cite{Ref:Kumar09,Ref:Mather12}.

\subsection{Smoothed Particle Hydrodynamics \label{subsec:SPH}}
Smoothed Particle Hydrodynamics is used for density-based control of a swarm to achieve spatially-organizing behaviors  \cite{Ref:MKumar11,Ref:Kumar13,Ref:Erkmen07}. The work in \cite{Ref:Michael08,Ref:Michael09} adopts a closely related  approach that abstracts the state of a formation of agents as a continuum parameterized by the formation position, orientation, and shape.

\subsection{Optimal Transport based Algorithm \label{subsec:optimal-transport}}
Optimal transport \cite{Ref:Villani08} allows to achieve spatially-organizing behavior by transporting a swarm of agents, represented as a probability distribution, from its current distribution to a desired distribution while optimizing a cost function \cite{Ref:Bandyopadhyay14MSC,Ref:Bandyopadhyay17_Rainbow,Ref:De2018optimal}.

\section{Centralized Optimization Algorithms}
\label{sec:centralized-optimization-algo}
Centralized optimization algorithms solve multi-agent optimization problems by collecting states and information from all agents, computing the optimal solution based on this information, and assigning the relevant outputs to each agent. A wealth of centralized optimization algorithms are available to solve multi-agent tasks ranging from pattern formation and coverage to task allocation and motion planning.

\subsection{Mixed-Integer Linear Programming (MILP) and Mixed-Integer Nonlinear Programming (MINLP) Algorithms \label{subsec:MILP}}
\textbf{\textit{Mathematical description and guarantees:}}
A mixed-integer linear program has the form
\begin{subequations}
\label{eq:MILP}
\begin{align}
\underset{x}{\text{minimize }} & f \cdot x \label{eq:MILP:cost}\\
\text{subject to } & A \cdot x = b \label{eq:MILP:equalityc}\\
& x^i\in \mathbb{Z} \quad \forall i \in \mathcal{Z}\\
& x^i\in \mathbb{R} \quad \forall i \not \in \mathcal{Z} 
\end{align}
\end{subequations}
where $x=[x^1, \ldots, x^{|\mathcal{V}|}]$ is the concatenation of the agents' optimization variables, $f$ and $b$ are vectors, $A$ is a matrix, and the set $\mathcal{Z}$ denotes the subset of the agents' optimization variables that are constrained to be integral.
Mixed-integer linear programs (MILPs) can be used for optimal control of systems with linear system dynamics, linear constraints, and complex \emph{logical} constraints encoded by integer-valued variables. Mixed-integer quadratic programs (MIQPs) and mixed-integer convex programs (MICPs) allow further expressivity by incorporating quadratic and convex constraints. We refer the reader to \cite{Ref:Morari99} for a comprehensive introduction to MILPs and MIQPs for control of dynamical systems.

MILPs are NP-hard in the general case; however, a number of efficient numerical solvers are available that can find solutions to problems with hundreds of thousands of variables in minutes to hours \cite{Mittelmann2016}.
Scalability of MILPs with the number of agents is problematic: in general, the complexity of MILPs is exponential in the problem size (and, in particular, in the number of agents). For instance, in \cite{Ref:Kushleyev13}, the authors demonstrate path planning and collision avoidance on a hardware testbed with 20 UAVs: however, the resulting controller requires several seconds to produce a feasible solution, and multiple minutes to achieve an optimal solution.

In general, MILPs are not amenable to efficient distributed solution.
However, linear relaxation of MILPs (LPs) are amenable to a distributed implementation (discussed in the next section); rounding can then be used to recover a suboptimal integral solution \cite{Ref:Zhe13,Ref:Shamma05}.

MILPs can also be solved in a shared-world framework (discussed in the \nameref{sidebar:shared-world-optimization} sidebar): in \cite{Ref:Alighanbari05}, the authors proposed MILP models that are robust to synchronization errors among the agents, and the approach is demonstrated in hardware in \cite{Ref:Bethke06}. %

\textbf{\textit{Applications:}}
MILPs and MICPs have been widely used for task allocation and path planning in multi-agent robotic systems problems. In particular, MILPs can be used to solve \emph{simultaneous} task allocation and path planning problems with applications including tracking of multiple targets \cite{Ref:Zhe13},  task assignment for UAVs \cite{Ref:How03,Ref:Alighanbari05}, spacecraft formation flying \cite{Ref:How02,Ref:Tillerson02}, and capture-the-flag games (RoboFlag \cite{Ref:Earl02modeling,Ref:Shamma05}). 
MILPs can encode collision avoidance, communication, and connectivity constraints \cite{Ref:How02,Ref:Richards2002aircraft,Ref:Atay06,Ref:How06,Ref:Bezzo11,Ref:Kuwata11}.%

\subsection{Linear and Convex Optimization\label{subsec:centralized-LP}}
\textbf{\textit{Mathematical description and mathematical guarantees:}}
A convex program has the form
\begin{subequations}
\label{eq:centralizedCP}
\begin{align}
\underset{x}{\text{minimize }} & g(x) \label{eq:centralizedCP:cost}\\
\text{subject to } & A x = b \label{eq:centralizedCP:equalityc}\\
& f(x)\leq 0 \label{eq:centralizedCP:inequalityc}
\end{align}
\end{subequations}

where the cost function $g(\cdot)$ and the inequality constraints $f(\cdot)$ are convex functions. In linear programs, $g(\cdot)$ and $f(\cdot)$ are affine.

Convex programs can be solved in time polynomial in the size of their inputs by interior point algorithms \cite{Karmarkar1984,Ref:Boyd09}. The simplex algorithm \cite{Dantzig1998} can also be used to solve linear programs very efficiently
for a wide class of problems of practical interest. 
State-of-the-art solvers can efficiently solve linear and convex programming problems with millions of variables: thus, problems with hundreds to thousands of agents can be solved efficiently \cite{Mittelmann2016}.

\textbf{\textit{Applications:}}
A number of variations of the  task allocation problem, a central problem in multi-agent robotics, can be posed as a linear program \cite{Ref:Bertsekas98b}, and can therefore be solved very efficiently. This observation has motivated significant research on extensions of the task allocation problem that capture additional constraints (in particular collision avoidance constraints) while preserving the tractability of the problem. 
In \cite{Ref:Acikmese06}, an algorithm for task allocation with collision avoidance constraints is proposed  for reconfiguration of spacecraft formations: a heuristic is proposed to approximate the collision avoidance constraints with a convex constraint. %
 \cite{Ref:Kumar14} shows that, if a cost function minimizing the integral of the squared velocity of the agents is selected, the task allocation problem is guaranteed to yield collision-free trajectories; under mild conditions, the result also holds for agents of finite size. The authors also propose a distributed implementation of the algorithm, based on message-passing, that yields locally-optimal and collision-free solutions. 
Convex optimization can also be used to approximately solve combinatorial sensor selection placement problems for target tracking \cite{Ref:Joshi2009sensor}.

\subsection{Markov Decision Processes (MDP) and Partially Observable Markov Decision Processes (POMDP) \label{subsec:MDP}}
\textbf{\textit{Mathematical description and mathematical guarantees:}}
A Markov decision process (MDP) is described by the tuple $(S,A_s,P_a,R_a)$, where $S$ is the set of possible states of the system, $A_s$ is the set of actions that can be performed in state $s\in S$, $P_a(s,s')$ is the probability of transitioning to state $s'$ from state $s$ if action $a$ is taken, and $R_a(s,s')$ is the immediate reward associated with transitioning from state $s$ to state $s'$ by taking action $a$. The key assumption underlying MDPs is that transitions and rewards obey the Markov property, i.e. they only depend on the current state and not on the prior trajectory followed by the system.
The goal of a Markov decision process problem is to select the \emph{policy} $\pi: S\mapsto A_s$ that maximizes the discounted reward received by the system over an infinite horizon. Formally, a Markov decision process can be stated as
\[
\underset{\pi}{\text{minimize}} \sum_{t=0}^\infty \gamma^t R_{\pi(s_t)}(s_t,s'_t) 
\]
where $\gamma\in(0,1]$ is denoted as the discount factor.

In Partially Observable Markov Decision Processes (POMDPs), agents can not directly observe the state of the system; rather, they have access to noisy observations of the state and must plan according to their \emph{belief} state, i.e. a probability distribution over the possible states where the system may be.

In general, the state size (and therefore the computational complexity) of multi-agent MDPs scales exponentially with the number of agents; thus, MDPs with more than a few agents are often intractable.
Furthermore, the computational complexity of POMDPs scales exponentially with the state size.
A number of techniques are available to make large-scale MDPs and POMDPs tractable: in particular, \cite{Ref:Chen14} proposes hierarchical decomposition techniques for multi-agent MDPs. We refer the reader to \cite{Ref:Amato13} for a taxonomy of multi-agent POMDPs and a survey of solution algorithms.

\begin{figure}
\centering
\includegraphics[width=\columnwidth]{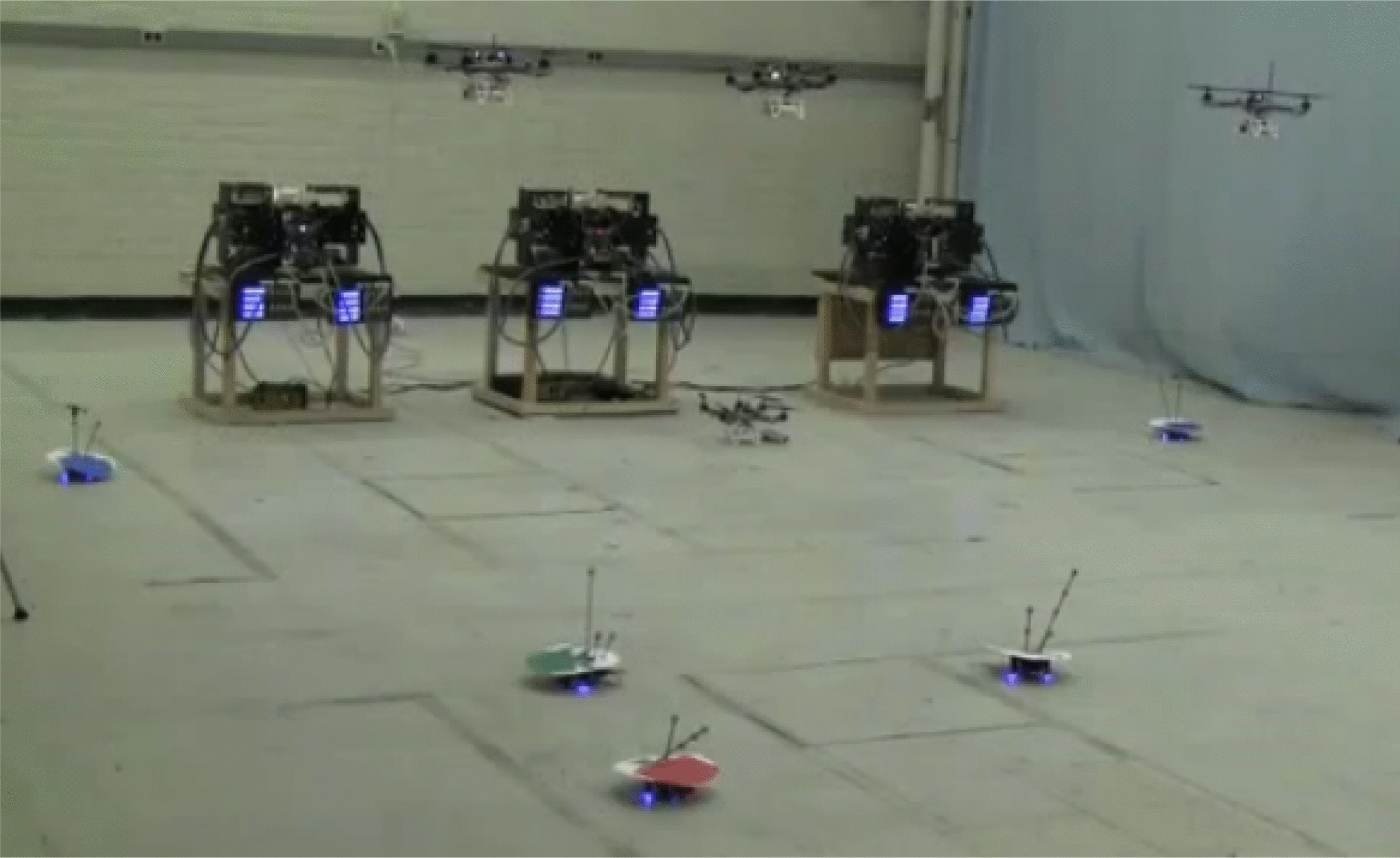}
\caption{
A team of three UAVs and a ground vehicle performs a persistent search and track mission using a Markov Decision Process, a \textbf{centralized optimization} algorithm.
Centralized optimization algorithms solve multi-agent optimization problems by collecting states and information from all agents, computing the optimal solution based on this information, and assigning the relevant outputs to each agent.
In this example, the agents fly to a surveillance region, find targets of interest, and track them. The base station is out of line of sight: a vehicle is autonomously deputized as a communication relay to ensure that a communication link is maintained. The battery rate of discharge are partially unknown: the system uses a probabilistic model for battery health, which is refined during the mission based on the agents' observations.
The three UAVs and the blue ground vehicle use a centralized MDP formulation to solve the persistent search and track perform with connectivity and battery charge constraints. The red and green ground vehicles are targets; the white ground vehicle is neutral.
\copyright 2014 IEEE. Reprinted, with permission, from \cite{Ref:Ure15}.
}
\label{fig:centralized:mdp:batteryswap}
\end{figure}

\textbf{\textit{Applications:}}
MDPs \cite{Ref:Puterman14} and POMDPs \cite{Ref:Monahan82} offer a natural description of multi-agent systems that captures the stochastic nature of the agents' dynamics and of the environment. Furthermore, MDPs can be leveraged to compute optimal \emph{policies} offline for each state of the system: such policies allow systems of agents to react to stochastic occurrences in real-time with minimal computation and communication overhead. As such, MDPs are well-suited to path planning under uncertainty (especially in the discrete domain) and task allocation applications for multi-agent systems. Critically, for multi-agent systems, MDPs can explicitly model the agents' \emph{coordination mechanism} \cite{Ref:Boutilier99b} and capture the effect of miscoordination on the system. Solution algorithms for POMDPs that result in \emph{distributed} online policies and model inter-agent communication are proposed in \cite{Ref:Goldman04}. The work in \cite{Ref:Ali15} proposes techniques for multi-agent path planning using Partially Observable Markov Decision Processes (POMDPs). In \cite{Ref:Ure15}, the authors leverage MDPs to schedule battery swaps in a hardware platform with three UAVs. 
\revXXIa{
Figure \ref{fig:centralized:mdp:batteryswap} (from \cite{Ref:Ure15}) shows the team of three drones and a ground vehicle continuously monitoring  ground targets; the system performs monitoring, maintains network connectivity, and schedules recharging events using a Markov Decision Process (MDP) that includes a probabilistic model of battery health.
}

\subsection{Multi-Agent Traveling Salesman Problem (TSP) \label{subsec:TSP}}

\textbf{\textit{Mathematical description and mathematical guarantees:}}
The traveling salesman problem (TSP) consists of finding a route of minimal length that visits a set of prescribed points. The closely related Orienteering problem consists of finding a route that maximizes the number of points visited without exceeding a path length constraint.  
The TSP and the orienteering problem are not algorithms but rather combinatorial problems, both NP-hard. However, a number of high-quality approximation algorithms are available to solve many variations of the TSP and the Orienteering problem, including the multi-agent TSP and the team orienteering problem. In turn, solutions to the TSP and Orienteering problems are often used as fundamental building blocks to solve complex problems in robotics. 
\cite{Ref:Betkas06} provides an excellent review of formulations and algorithms for the multi-agent Traveling Salesman Problem (m-TSP). \cite{Ref:Vansteenwegen11} provides a similar review for the Orienteering and Team Orienteering problems. 
In \emph{dynamic} versions of the TSP, the list of locations to be visited is progressively revealed to the agents and drawn from a stochastic distribution. The work in \cite{Ref:Bullo11} provides solutions to the multi-agent Dynamic Traveling Repairman Problem, a version of the multi-agent TSP where agents must spend a set amount of time at each visited location. 

\textbf{\textit{Applications:}}
The m-vehicle Traveling Salesman Problem and the dynamic m-TSP can be used to solve spatial task allocation problems \cite{Ref:Bullo11}; the team orienteering problem can be used as a building block for persistent monitoring and information-gathering problems \cite{yu2016correlated}.

\subsection{Coordinated Combinatorial and Sampling-based Motion-Planning Algorithms \label{subsec:combinatorial-mp}}

Combinatorial and sampling-based motion planning algorithms for single-agent systems \cite{Ref:Lavalle06} can be readily extended to multi-agent systems.
The workspace of the multi-agent system is defined as the union of the individual agents' workspaces; each point in the workspace identifies the configuration of \emph{all} agents in the system.

In \emph{combinatorial} algorithms, the free workspace (i.e. the set of points in the workspace where no agent is in collision with another agent or with the environment) is discretized and represented as a graph, with vertices corresponding to points or regions in the workspace and arcs modeling the agent's ability to move from one region to the next; the motion planning problem is thus reduced to a graph search problem, which can be solved by algorithms such as Dijkstra's algorithm \cite{Dijkstra1959} and A* \cite{HartNilssonEtAl1968}.
The resulting problem is generally NP-hard \cite{Ref:LaValle13}.
Nevertheless, several extensions of combinatorial algorithms to cooperative motion planning in multi-agent systems have been proposed;  \cite{Ref:Sturtevant15} has a thorough review.
In particular, \cite{Ref:Ryan08} and \cite{Ref:Standley10} propose techniques to efficiently prune the search space, and an anytime algorithm is proposed in \cite{Ref:Standley11}.
In \cite{Ref:Sharon13}, the authors propose a search technique based on a two-level search
to quickly prune suboptimal solutions; %
\cite{Ref:Sturtevant15} uses %
a "conflict tree" to keep track of conflicts between the agents' paths.
 
 In \emph{sampling-based} motion planning algorithms, a graph representation of the free space is built by sampling points in the free workspace and connecting them if a collision-free route exists between them. A variety of strategies for sampling and connecting samples exist: we refer the reader to \cite{Ref:Lavalle06} for a review.
The multi-agent motion planning problem is generally PSPACE-complete
\cite{Hopcroft1984,Solovey2016a,Hearn2005}: thus, the complexity of sampling-based multi-agent motion planning algorithms  also grows exponentially with the number of agents. 
Research in the field has focused on the identification of efficient data structures to represent and sample the joint state space of the agents. 
 Several sampling-based motion planning algorithms have been proposed for multi-agent systems \cite{Sanchez2002,Wagner2015,Solovey2016,Dobson2017}; 
 in particular, \cite{Wagner2015,Dobson2017} offer optimality guarantees.  A key problem in sampling-based motion planning is to efficiently assess the ``distance'' between sampled points - 
the work in \cite{Atias2017} proposes and assesses the performance of different metrics to efficiently estimate the ``distance'' between samples in the multi-agent setting.

\subsection{Dynamic inversion and differential flatness\label{subsec:differential-flatness}}

In differentially flat systems \cite{FliessLevineEtAl1995}, the control inputs required to track a given state trajectory can be computed analytically by inverting the system dynamics. 
Many dynamical systems of interest (in particular, quadcopters and groups of quadcopters lifting a payload \cite{SreenathKumar2013}) enjoy the differentially flatness property \cite{MurrayRathinamEtAl1995}. Accordingly, control techniques based on differential flatness and dynamic inversion have proven to be very popular for cooperative lifting applications where multiple quadrotors are attached to a payload rigidly \cite{Mellinger2013} or  through flexible cables \cite{FinkMichaelEtAl2010,MichaelFinkEtAl2011}.
Algorithms in this class are generally used to compute a sequence of dynamically feasible poses that the agents should traverse; low-level tracking controllers and simple collision avoidance heuristics are then used to guide the agents  between the poses.

\subsection{Direct Methods for Optimal Control \label{subsec:direct-methods}}
Direct methods for trajectory optimization \cite{Ref:VonStryk92} %
can be used to design multi-agent trajectories that approximately minimize a cost function over a known environment. Applications of direct methods for trajectory optimization to multi-agent robotic systems include area coverage, goal searching, and distributed estimation. While direct methods are generally not suitable for real-time optimization due to their high computational complexity, they can be used to design open-loop trajectories that can be tracked in real-time by a lower-level controller.
Direct methods have been employed as part of larger control architectures and are technologically mature. For instance, in \cite{Ref:Leonard07b}, the authors leverage direct methods to design trajectories for autonomous underwater vehicles that maximize the information gain over an information field with known distribution; such techniques are demonstrated in the field with a system of six underwater gliders in \cite{PaleyZhangEtAl2008,Ref:Leonard10b}.

\subsection{Multi-agent Reinforcement Learning algorithms}
\label{subsec:RL}
Multi-agent reinforcement learning (MARL) enables robots to adapt to unknown environments and learn high-quality individual and joint policies with little to no \textit{a priori} information. In \emph{independent learners} MARL algorithms, each agent attempts to learn an optimal policy for itself; in \emph{joint action learners} algorithms, agents collaborate to learn a \emph{joint} optimal policy. We refer the reader to \cite{Ref:Boutilier98} and \cite{Ref:Mataric1997reinforcement} for a taxonomy of problems in multi-agent reinforcement learning and to \cite{Ref:Busoniu08,zhang2019MARL} for more recent reviews of the problem and of solution algorithms. 
\cite{Ref:Chalkiadakis03} proposes a Bayesian multi-agent reinforcement learning algorithm that explicitly captures the value of information sharing between the agents and the cost of exploration. In the model, each agent holds a Bayesian model of the other agents' policy, allowing agents to adapt to each other's policies with no communication.
The work in \cite{Ref:Liu16} proposes a centralized hierarchical reinforcement learning approach for human-in-the-loop control of robots in urban search-and-rescue situations: reinforcement learning is used for multi-robot exploration and task allocation on a two-robot hardware testbed.

A recent line of work explores the role of \emph{locality} (intuitively, the property that an agent's state and decisions only influence the rewards of nearby agents) on multi-robot systems, and leverages locality to significantly improve the scalability of MARL algorithms for problems exhibiting a high degree of locality \cite{pmlr-v120-qu20a,Lin2020MARL,Qu2020MARL}.

\subsubsection{Multi-Armed Bandits \label{subsec:MAB}} Multi-Armed Bandits (MAB) problems  \cite{Ref:Gittins79} are a class of reinforcement learning problems where an agent must repeatedly select an action from a finite set of possibilities. Each action yields a stochastic reward with an initially unknown distribution; the goal is to maximize the expected discounted reward collected by the agent over an infinite horizon. Variations of the MAB problem allow the stochastic distribution of the rewards to evolve in time, possibly depending on whether the action is selected or not \cite{Ref:Whittle88}. In multi-agent MAB problems, multiple agents may each select an action independently.  MAB problems capture well the trade-off between exploration and exploitation: as such, they have been employed for task allocation \cite{Ref:Le06,Ref:Le08} and surveillance applications \cite{Ref:Leonard16}%
. MAB algorithms can also be implemented in a shared-world framework \cite{Ref:Leonard16}; a consensus algorithm is used to synchronize the agents' beliefs.

\subsection{Frontier Techniques \label{subsec:centralized-frontier}}
Frontier techniques for area exploration leverage a cost-based heuristic to assign robots to unexplored regions of the environment so as to approximately minimize the time required for full exploration \cite{Ref:Yamauchi98,Ref:Simmons2000coordination,Ref:Burgard05}. Frontier techniques are generally centralized; however, they can be adapted to accommodate the agents' limited communication range with good performance in practical applications \cite{Ref:Burgard05}.
Frontier techniques have seen wide adoption as part of robotic stacks for applications including urban search-and-rescue and reconnaissance (\cite{Ref:Olson12}, with a fourteen-robot team), and proposed Lunar sample collection missions (\cite{Ref:Eich14}, with a shared-world three-robot architecture).

\subsection{Network Flow Algorithms \label{subsec:network-flow}}
In network flow algorithms, the environment that the agents move in is represented as a capacitated graph.
The agents move from origin nodes to destination nodes; the assignment of robotic agents to destinations may either be fixed or an optimization variable. A traversal cost is associated with each edge; the agents' routes are optimized so as to minimize a cost function defined on the edges while satisfying constraints (e.g., capacity constraints on the edges).

Network flow formulations have been proposed for Air Traffic Control \cite{Ref:Menon04} and for control of  fleets of self-driving vehicles offering on-demand transportation (also known as Autonomous Mobility-on-Demand) \cite{Ref:Pavone11b, RossiIglesiasEtAl2018, RossiZhangEtAl2017}. Network flow problems can be solved either as linear programs or with dedicated combinatorial algorithms \cite{Ref:Goldberg90}: problems with thousands of robotic agents can be solved efficiently on commodity hardware.

\section{Distributed Optimization Algorithms}
\label{sec:distributed-optimization-algorithms}
While all the centralized optimization algorithms presented in the previous section can be implemented in a distributed manner through a shared-world approach (discussed in the \nameref{sidebar:shared-world-optimization} sidebar), concerns about robustness and bandwidth requirements have motivated the development of a number of \emph{distributed} optimization algorithms that allow agents to jointly solve optimization problems through limited information exchange and local computation. 

\subsection{Distributed Linear and Convex Programming \label{subsec:LP}}

\textbf{\textit{Mathematical description:}} Distributed linear and convex programming algorithms enable a set of agents to collectively solve a convex program in the form 

\begin{subequations}
\label{eq:convexprogram}
\begin{align}
\text{minimize } & \sum_{i=1}^{|\mathcal V|} g^i(x^i) \label{eq:convexprogram:cost}\\
\text{subject to } & h(x^1, \ldots, x^{|\mathcal V|}) = 0 \label{eq:convexprogram:equalityc}\\
& f(x^1, \ldots, x^{|\mathcal V|})\leq 0 \label{eq:convexprogram:inequalityc}
\end{align}
\end{subequations}
where the cost function \eqref{eq:convexprogram:cost} is convex and the constraints are expressed as either affine equalities \eqref{eq:convexprogram:equalityc} or convex inequalities \eqref{eq:convexprogram:inequalityc}.
Each agent is responsible for determining its own variables $x^i$; the goal function is the sum of the agents' individual goals, and the constraints introduce a \emph{coupling} between the agents' decisions.

The dual decomposition algorithm \cite[Ch. 6]{Bertsekas1999}, subgradient methods  \cite{Ref:Nedic09,Ref:Nedic2009TAC,Ref:Nedic2015opt}, and the Alternating Direction Method of Multipliers (ADMM) algorithm \cite{GabayMercier1976,GlowinskiMarroco1975}  can efficiently solve problems in the form of \eqref{eq:convexprogram} in presence of a fully-connected network communication topology; We refer the reader to the monography \cite{Ref:Boyd11} for a comprehensive overview. 

Several distributed algorithms tailored to \emph{linear} programs are also available. The work in \cite{Ref:Cortes15} proposes a distributed linear programming algorithm that is robust with respect to finite disturbances and to link failures. %
  The paper \cite{Ref:Bullo12} devises a distributed simplex algorithm to solve degenerate linear programs on time-varying communication topologies that is tailored to distributed task allocation. A distributed algorithms to solve abstract linear programs \cite{AgarwalSen2001}, a generalization of linear programs, is proposed in \cite{Ref:Bullo07}.
We refer the reader to \cite{Ref:Yang10} for a cohesive survey of distributed linear and convex optimization algorithms.

\textbf{\textit{Mathematical guarantees:}}
ADMM algorithms are guaranteed to converge on static communication networks under very mild assumptions on the shape of the cost function \eqref{eq:convexprogram:cost}. The algorithm in \cite{Ref:Cortes15} is also guaranteed to converge and, in addition, it offers robustness guarantees. Finally, the techniques in \cite{Ref:Bullo07} and \cite{Ref:Bullo12} are amenable to implementation on time-varying communication topologies.

\textbf{\textit{Communication bandwidth:}}
In \cite{Ref:Cortes15}, messages of constant size are exchanged on every edge of the communication graph. Conversely, in the absence of a central coordinating authority, the dual variable update of the ADMM algorithm requires agents to communicate their local variables to \emph{all} other agents through multi-hop communication.

\textbf{\textit{Applications:}} Applications of distributed linear and convex programming include distributed estimation \cite{Ref:Boyd11}, task allocation \cite{Ref:Bullo12}, and formation control \cite{Ref:Bullo07}.

\subsection{Distributed Auction and Market-Based Techniques \label{subsec:auction}} 
\begin{figure}
\centering
\includegraphics[width=.5\textwidth]{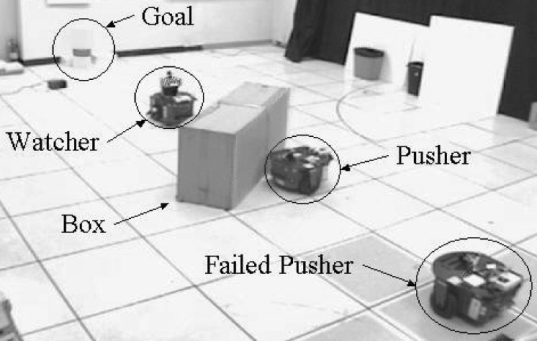}
\caption{
A heterogeneous team of robots performs a box-pushing task using the the MURDOCH auction-based task allocation system, an example of \textbf{distributed optimization}.
Distributed optimization algorithms allow agents to jointly solve optimization problems through information exchange and local computation, without delegating an individual agent to represent or solve the overall problem.
In the figure, two "pusher" robots push a box towards a goal. While pushing, the robots cannot see the goal: a "watcher" robot aids them by tracking the position of the box relative to the goal.
If a pusher robot fails, the remaining robot is left to push on its own. Once the failed robot is revived, it reenters the team, and all robots complete the task together. 
\copyright 2002 IEEE. Reprinted, with permission, from \cite{Ref:Gerkey02}.
}
\label{fig:auction:pushing}
\end{figure}

Market-based protocols for distributed task allocation have enjoyed immense success in the robotics community. In particular, 
auction-based protocols \cite{Ref:Bertsekas98,Ref:Pappas08,Ref:Shoham08,Ref:Gerkey02,Ref:Stojmenovic10,REf:Sujit11} are widely used for distributed coordination of agents.  The papers \cite{Ref:Dias99,Ref:Dias04,Ref:Arslan07,Ref:Marden2009cooperative} propose mechanisms  to obtain emergent coordination behavior from a network of self-interested agents, and \cite{Ref:Shehory98} proposes algorithms for distributed coalition formation for task allocation. 
\revXXIa{
Figure \ref{fig:auction:pushing} (from \cite{Ref:Gerkey02}) shows a heterogeneous team of robots performing a box-pushing task using an auction-based task allocation system and autonomously recovering from failures of individual robots.
}

\textbf{\textit{Mathematical description and guarantees:}} 
Consider a prototypical distributed auction algorithm with $n$ agents and $m \geq n$ tasks.
Let $\beta[i,\ell] \in \mathbb{R}$ the benefit of assigning task $\ell, \forall \ell \in \{1,\ldots,m\}$ to agent $i, \forall i \in \{1,\ldots,n\}$.
The $i^{\text{th}}$ agent keeps track of its bid $\alpha^i_k \in \{1,\ldots,m\}$. 
The $i^{\text{th}}$ agent maintains a local copy of the highest task prices $P^i_k[\ell]$ and highest bidder $B^i_k[\ell]$ for all $\ell \in \{1,\ldots,m\}$, which it updates as follows \cite{Ref:Pappas08}:
\begin{align}
P^i_{k+1}[\ell] &= \max_{j \in \mathcal{J}_k^i} P^j_{k}[\ell] \thinspace , \\
B^i_{k+1}[\ell] &= \max_{q \in \arg \max_{j \in \mathcal{J}_k^i} P^j_{k}[\ell]} B^q_{k}[\ell] \thinspace .
\end{align}
If the $i^{\text{th}}$ agent has been outbid, then it updates its bid and sets a new price for its updated bid as follows:
\begin{align}
\text{\textbf{if} }& P^i_{k+1}[\alpha^i_k] \geq  P^i_{k}[\alpha^i_k] \text{ and } B^i_{k+1}[\alpha^i_k] \neq i \text{, \textbf{then}} \nonumber \\
& \alpha^i_{k+1} = {\arg \max}_{\ell \in \{1,\ldots,m\}} (\beta[i,\ell] - P^i_{k+1}[\ell]) \thinspace, \\
& B^i_{k+1}[\alpha^i_{k+1}] = i \thinspace, \\
\ifonecolumn
& P^i_{k+1}[\alpha^i_{k+1}] = P^i_{k+1}[\alpha^i_{k+1}] + (\beta[i,\alpha^i_{k+1}] - P^i_{k+1}[\alpha^i_{k+1}]) \nonumber \\
& \qquad - \max_{\ell \in \{1,\ldots,m\} \setminus \{\alpha^i_{k+1}\}} (\beta[i,\ell] - P^i_{k+1}[\ell]) + \varepsilon \\
\else
& P^i_{k+1}[\alpha^i_{k+1}] =  \\
& \qquad P^i_{k+1}[\alpha^i_{k+1}] + (\beta[i,\alpha^i_{k+1}] - P^i_{k+1}[\alpha^i_{k+1}]) \nonumber \\
& \qquad - \max_{\ell \in \{1,\ldots,m\} \setminus \{\alpha^i_{k+1}\}} (\beta[i,\ell] - P^i_{k+1}[\ell]) + \varepsilon \nonumber \\
\fi
\text{\textbf{else} }& \alpha^i_{k+1} = \alpha^i_k \thinspace .
\end{align}
It can be shown that the above algorithm reaches equilibrium assignment with $\varepsilon$ error \cite{Ref:Pappas08}. 

\textbf{\textit{Communication bandwidth:}} The agents exchange pricing information with all other agents using multi-hop communication, resulting in demanding communication requirements.%

\textbf{\textit{Applications:}} \textit{Cooperative decision making:} cooperative transport with UAVs \cite{Ref:Maza11}, pattern formation \cite{Ref:Siegwart12}, network connectivity maintenance \cite{Ref:Zavlanos08}.

\subsection{Distributed Sequential Convex Programming \label{subsec:SCP}}
Non-convex multi-agent optimization problems can be iteratively solved using distributed sequential convex programming (SCP). In distributed SCP, the non-convex elements of the original optimization problem (e.g., cost function and collision avoidance constraints) are convexified using local function approximations.
The resulting convex problem is solved in a distributed fashion by leveraging the distributed convex programming algorithms discussed in the \emph{Distributed Linear and Convex Programming} section. 
The solution of the convex surrogate problem is then used for convexification in the next SCP iteration. 
Distributed implementations of SCP has been used for control of spacecraft swarms \cite{Ref:Morgan14,Ref:Morgan15_SATO} and motion planning \cite{Ref:Bandyopadhyay17_MA_SESCP,Ref:Bandyopadhyay17_MAMO_SESCP}.

\subsection{Distributed Dynamic Programming \label{subsec:distributed-dp}}
In distributed dynamic programming  algorithms \cite{Ref:Sukhatme04,Ref:Balch07},  each agent computes a partial solution to a dynamic programming problem based on its and its neighbors' information; agents iteratively communicate with their neighbors until convergence to the globally optimal solution.

Distributed dynamic programming has been used for task allocation \cite{Ref:Sukhatme04} and path planning \cite{Ref:Balch07} in multi-agent systems operating in environments equipped with a pre-deployed, static sensor network.

\subsection[Cooperative Model Predictive Control]{Cooperative Model Predictive Control \label{subsec:DecMPC}}

In cooperative Model Predictive Control (MPC) algorithms, agents repeatedly cooperate to solve a finite-horizon version of a global optimization problem, then implement the first step of the computed control policy, and repeat the optimization. Model-predictive control allows agents to incorporate new information about the system as they learn it;  the computational complexity of the optimization problem can be kept in check by repeatedly solving a version of the problem with a short horizon.

The finite-horizon global optimization problem is solved in a distributed fashion by leveraging the  distributed linear, convex, or sequential convex programming algorithms presented in this section.

The stability and optimality of the distributed, cooperative MPC problem is studied in \cite{VenkatRawlingsEtAl2005}.

\section{Local Optimization Algorithms}
\label{sec:local-optimization-algo}
In local optimization algorithms, each agent solves an optimization problem; while the resulting behavior is not generally optimal for the entire system, favorable global properties such as collision avoidance \revXXIa{and convergence} can often be guaranteed, and near-optimal behavior can be achieved in specific applications.

\subsection{Distributed Model Predictive Control (DMPC) \label{subsec:MPC}}
In distributed model predictive control (DMPC)
each agent employs a local model-predictive control algorithms with the goal of optimizing its own cost function; inter-agent communication is used to coordinate the agents' plans. Distributed MPC has been used for flocking and motion planning \ifjournalv\cite{Ref:Dunbar02,Ref:How06,Ref:How07MPC,Ref:Bemporad10,Ref:Kuwata11,Ref:Zhan13}\else\cite{Ref:Dunbar02,Ref:How06}\fi. We refer the reader to the excellent review in \cite{Ref:Scattolini09} for a thorough discussion.

\subsection{Decoupled and Prioritized Motion Planning Algorithms \label{subsec:sampling-based-mp}}

The computational complexity of centralized sampling-based motion planning algorithms (discussed in the section on \emph{Coordinated Motion-Planning Algorithms}) grows exponentially with the number of agents. This has motivated research in local optimization algorithms where each agent optimizes its own trajectory (leveraging either combinatorial or sampling-based motion planning) and communication is used to ensure that no inter-agent collisions occur.

Local multi-agent motion planning algorithms can be divided in three classes: decopuled, prioritized, and roadmap-based.

\begin{figure}
\centering
\includegraphics[width=\ifonecolumn .5\textwidth\else\columnwidth\fi]{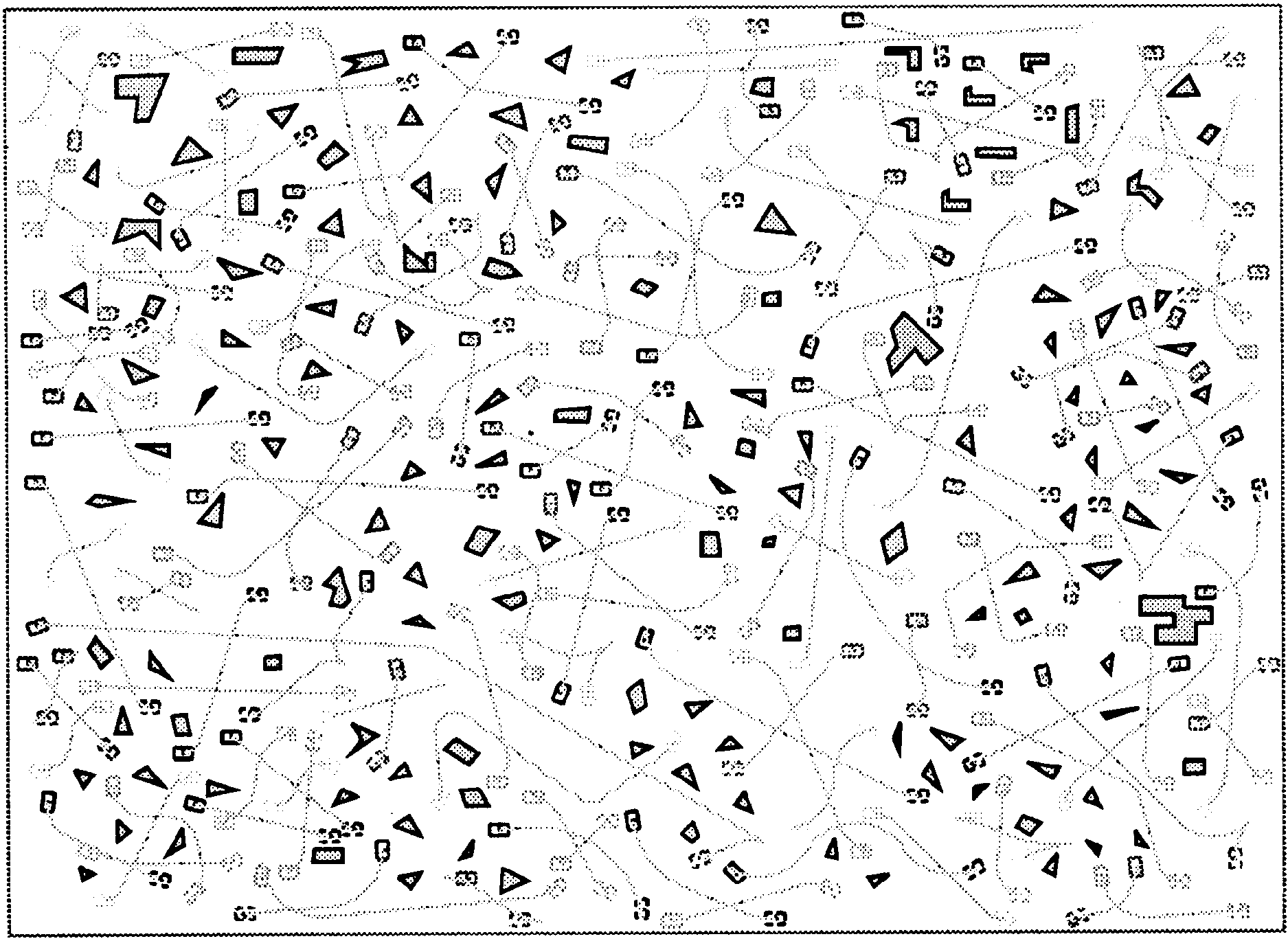}
\caption{
Path coordination problem, an example of \textbf{local optimization for global behavior}. A team of 150 robots finds independent, non-colliding paths in a crowded shared workspace by efficiently searching a \emph{coordination diagram} representing configurations where robot paths intersect.
In local optimization algorithms, each agent solves a local optimization problem to synthesize its control actions; while the resulting behavior is not generally optimal for the entire system, favorable global properties (e.g., convergence and collision avoidance) can be guaranteed.
\copyright 2002 IEEE. Reprinted, with permission, from \cite{Simeon2002}.
}
\label{fig:localopt:decoupled}
\end{figure}

In \emph{decoupled} algorithms, all agents plan their paths independently and simultaneously. Conflicts between the agents' computed trajectories are then resolved through a variety of techniques including velocity tuning (through path-velocity decomposition \cite{Kant1986} and bounding box representations \cite{Simeon2002}), generalized velocity obstacles \cite{Ref:Bareiss15}, and model-predictive control \cite{Ref:Bandyopadhyay17_MAMO_SESCP,Ref:Bandyopadhyay17_MA_SESCP}.
\revXXIa{
Figure \ref{fig:localopt:decoupled} (from \cite{Simeon2002}) shows a decoupled local multi-agent motion planning algorithm solving a path coordination problem with 150 mobile agents in a cluttered environment.
}

In \emph{prioritized} algorithms, agents take turns planning according to a priority scheme \cite{Erdmann1987,Bennewitz2002,Clark2002,Berg2005,Ref:Silver05,Ref:How12}; each agent considers the paths planned by higher-priority agents as dynamic obstacles. Selection of a priority scheme is critical to the performance of the algorithm; the priority scheme can be 
randomized \cite{Ref:Silver05}, use heuristics based on the distance to the target \cite{Berg2005} or the ``crowdedness'' of each robot's workspace \cite{Clark2002}, leverage merit-based token passing \cite{Ref:How12}, or be optimized at run-time with  randomized  hill-climbing \cite{Bennewitz2002}.

Finally, \emph{roadmap-based} approaches represent an intermediate solution between centralized and decoupled algorithms. In such algorithms, each agent individually computes a road-map of promising routes; agents then coordinate to select a set of collision-free trajectories from within their individual roadmaps \cite{LavalleHutchinson1998, Svestka1998,Ghrist2005}.

\subsection{Formal Methods and Game Theory \label{subsec:formal-methods}}
Formal methods and game-theoretical worst-case optimization are used in concert with low-level control primitives for multi-agent motion planning with guaranteed collision avoidance.
In \cite{TomlinPappasEtAl1998}, a  game-theoretic approach  is used to  compute  safe  sets  of  initial  conditions  and control  inputs in presence of other  agents (modeled as bounded disturbances). The approach is used to design provably-safe coordinated maneuvers for conflict resolution and collision avoidance between multiple aircraft.
In \cite{Ref:KressGazit08}, a similar technique is used to design probably-safe high-level controllers for self-driving vehicles.

\ifhtml
\else
\ifjournalv\else \vspace{-1.8mm} \fi
\section*{Conclusion}  \label{sec:conclusion}
\ifjournalv\else \vspace{-1.8mm} \fi
\fi

The proposed classification and the properties shown in Table \ref{tab:all_algos} highlight some surprising characteristics of collective behavior algorithms.
The majority of existing mathematical techniques are tailored to either low-level spatially organizing tasks (e.g., bio-inspired algorithms and density-based control) or high-level coordination applications (e.g., state machines and optimization-based algorithms).
\revXXIa{However, with the exception of Artificial Potential Functions, no single mathematical technique has been adapted to a wide variety of tasks that include both low-level and high-level application.}
This may motivate further research into non-APF algorithms for multi-agent systems that share APF's key properties of simplicity, scalability, and ability to incorporate a variety of constraints. %

Consensus, geometric algorithms, and APFs are often used as building blocks in more complex algorithms or autonomy stacks. 
For exmple, consensus is used as a coordination subroutine in a number of distributed optimization algorithms, and APFs and geometric algorithms are ubiquitously used for collision avoidance and tracking of trajectories produced by a higher-level controller (see, for instance, \cite{Ref:Leonard10b} and \cite{Ref:Egerstedt2001formation}). However, rigorously characterizing the end-to-end performance of such ``meta-algorithms'' remains an open challenge.

Relatively few algorithms are field-tested. %
Algorithms that have seen adoption in the field generally exchange very simple information (e.g., the agents' locations) or rely on centralized implementations. 
The scarcity of field-tested distributed algorithms might arise from the difficulty of characterizing and certifying the behavior of an entire multi-agent system when distributed algorithms are used. To overcome this, we believe that future work in design of collective behavior algorithms should prioritize (i) research in formal methods, theoretical analysis, and adoption of tools from the distributed algorithms literature to provide stronger guarantees for distributed systems and (ii) creation of standardized software and hardware testbeds to characterize the end-to-end behavior of such systems (the Robotarium remotely accessible multi-robot test-bed \cite{PickemGlotfelterEtAl2017,Egerstedt2020Robotarium} represents a development in this direction).

The classification proposed in this paper admits several promising extensions for future investigation. In particular, we hope to evaluate the performance of collective behavior algorithms according to additional metrics including 1) bandwidth use in broadcast and in point-to-point networks, 2) computational complexity, 3) availability of formal guarantees, 4) resilience to disruptions in communication network and to \emph{adversarial} failures, and 5) availability of a reference implementation.
We also wish to explore other possible classification for coordination algorithms based, for instance, on the content of messages exchanged by the agent (which vary from simple ``beacon'' messages reporting the agent's location to complex messages carrying intentions and bids), and the communication topology induced by the algorithm (single-hop vs. multi-hop).
Finally, we plan to further explore high-level multi-agent tasks, including simultaneous localization and mapping (SLAM) and adversarial ``swarm vs. swarm'' problems, and to assess the applicability and performance of collective behavior algorithms with respect to such applications.

\section*{Acknowledgments}
The authors would like to thank Daniel Friedman for the helpful comments on this work.
Part of this research was carried out at the Jet Propulsion Laboratory, California Institute of Technology, under a contract with the National Aeronautics and Space Administration (80NM0018D0004). Federico Rossi and Marco Pavone were partially supported by the Office of Naval Research, Science of Autonomy Program, under Contract N00014-15-1-2673. \textcopyright 2021. All rights reserved.

\bibliographystyle{IEEEtran}
{\small
\bibliography{SurveyBib}
}

\newpage
\ifonecolumn
\processdelayedfloats

\sidebars
\fi
\clearpage

\section[Algorithm Classification]{Sidebar: Classification of Collective Behavior Algorithms}
\label{sidebar:algorithm-classification}

\revXXIa{
We classify collective behavior coordination algorithms for multi-robot systems in ten broad classes, as shown in Figure \ref{fig:all_algos}. The classification process is synthesized in the flowchart in Figure \ref{fig:flowchart}. This sidebar presents a brief reference; a detailed description of each class is presented in the remainder of the paper.

\subsection{Consensus algorithms}
In consensus algorithms, each agent updates its own state as a weighted average of its neighbors' states. The set of neighbors can change in time based on the agents' motion. %

\subsection{Artificial Potential Functions (APF)}
APF algorithms synthesize agents' control inputs using the gradient of a suitably-defined potential function, where goals act as attractors, and obstacles have a repulsive effect. 

\subsection{Distributed Feedback Control}

In distributed feedback control algorithms, each agent is endowed with a classical feedback controller (i.e., the concatenation of an observer and a linear feedback controller)
whose input is the concatenation of the agent's output and the outputs of its neighbors. The overall system behaves as a dynamical feedback system where the sparsity pattern of the controller reflects the topology of the agent's communications.

\subsection{Geometric Algorithms}
In geometric algorithms, agents use their neighbors' location and speed information (often obtained through sensory information, with minimal communication) to make decisions and perform spatially organizing tasks and path planning.

\subsection{State Machines and Behavior Composition}

State machines and behavior composition algorithms rely on \emph{discrete} descriptions of the agent's states and actions, and design complex orchestrations of transitions and message exchanges to achieve a desired global behavior. Such algorithms are especially popular for combinatorial cooperative decision-making tasks such as task allocation. %

\subsection{Bio-Inspired Algorithms}
Bio-inspired algorithms mimic the behavior of swarms of animals such as insects and fish. 
They are often similar in spirit to state machines and behavior composition algorithms: their defining characteristics are (i) their biology-inspired heritage and (ii) the use of very limited communication and on-board computation resources, often resulting in low bandwidth use and good scalability.

\subsection{Density based Control}
Density-based algorithms  adopt an \emph{Eulerian} framework by treating agents as a continuum and devising  control laws to drive the agents' density towards a specified distribution.

\subsection{Centralized optimization algorithms}

Centralized optimization algorithms solve multi-agent optimization problems by collecting states and information from all agents, computing the optimal solution based on this information, and assigning the relevant outputs to each agent. A wealth of centralized optimization algorithms are available to solve multi-agent tasks ranging from pattern formation and coverage to task allocation and motion planning.

\subsection{Distributed Optimization Algorithms}
Distributed optimization algorithms allow agents to jointly solve optimization problems through information exchange and local computation, \emph{without} delegating an individual agent to represent or solve the overall problem.

\subsection{Local optimization algorithms for global behavior}

In local optimization algorithms, each agent solves a local optimization problem to synthesize its control actions; while the resulting behavior is not generally optimal for the entire system, favorable global properties (e.g., convergence and collision avoidance) can be guaranteed.
}

\clearpage

\section[Communication Structure]{Sidebar: Communication Structure of Collective Behavior Algorithms}
\label{sidebar:shared-world-optimization}
\begin{figure}[h!]
\centering

    \begin{subfigure}[b]{0.25\textwidth}
    \includegraphics[height=\textwidth]{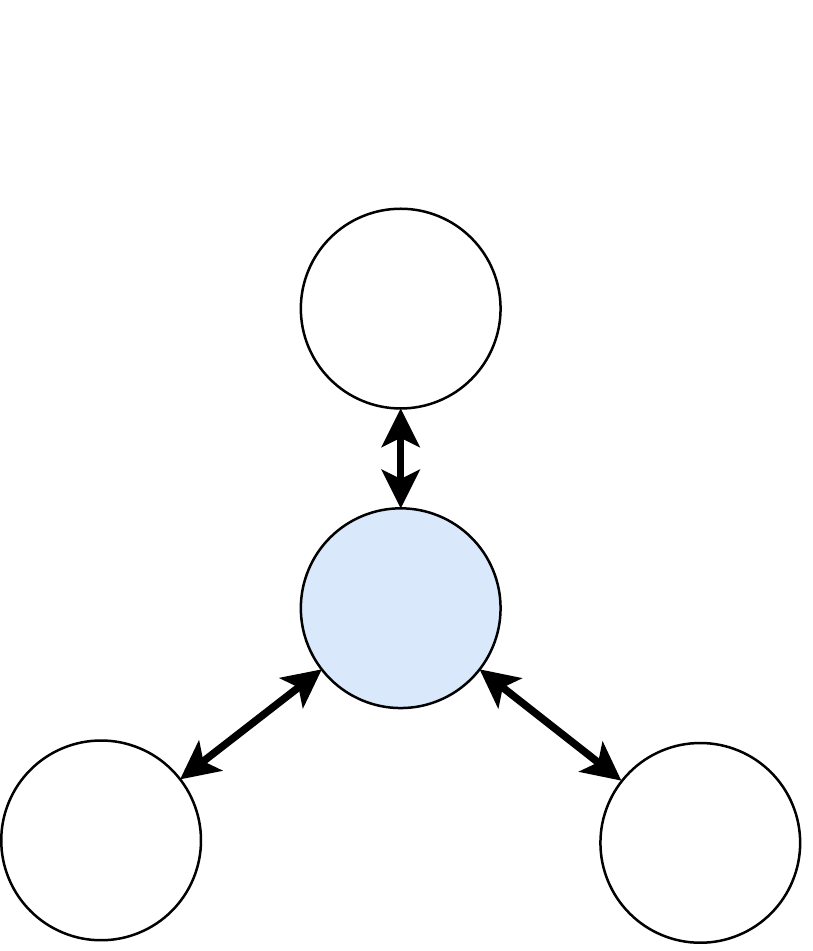}
    \caption{\label{fig:centralized}}
    \end{subfigure}%
        \begin{subfigure}[b]{0.25\textwidth}
    \includegraphics[height=\textwidth]{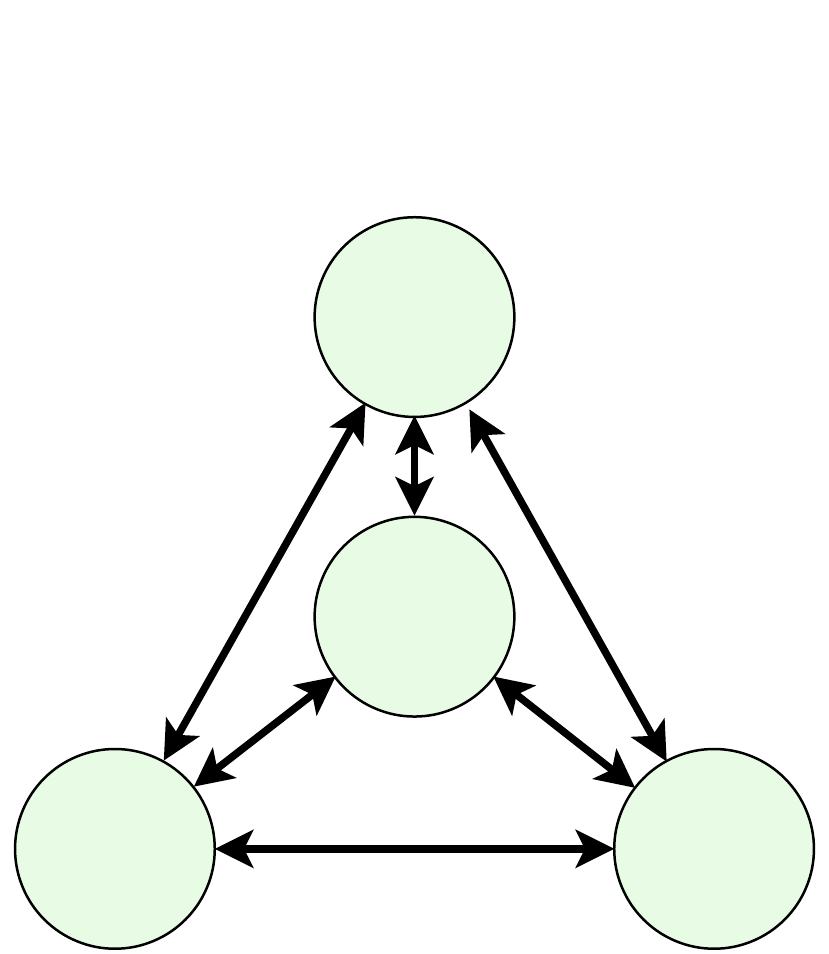}
    \caption{\label{fig:distributed}}
    \end{subfigure}\quad
        \begin{subfigure}[b]{0.25\textwidth}
    \includegraphics[height=\textwidth]{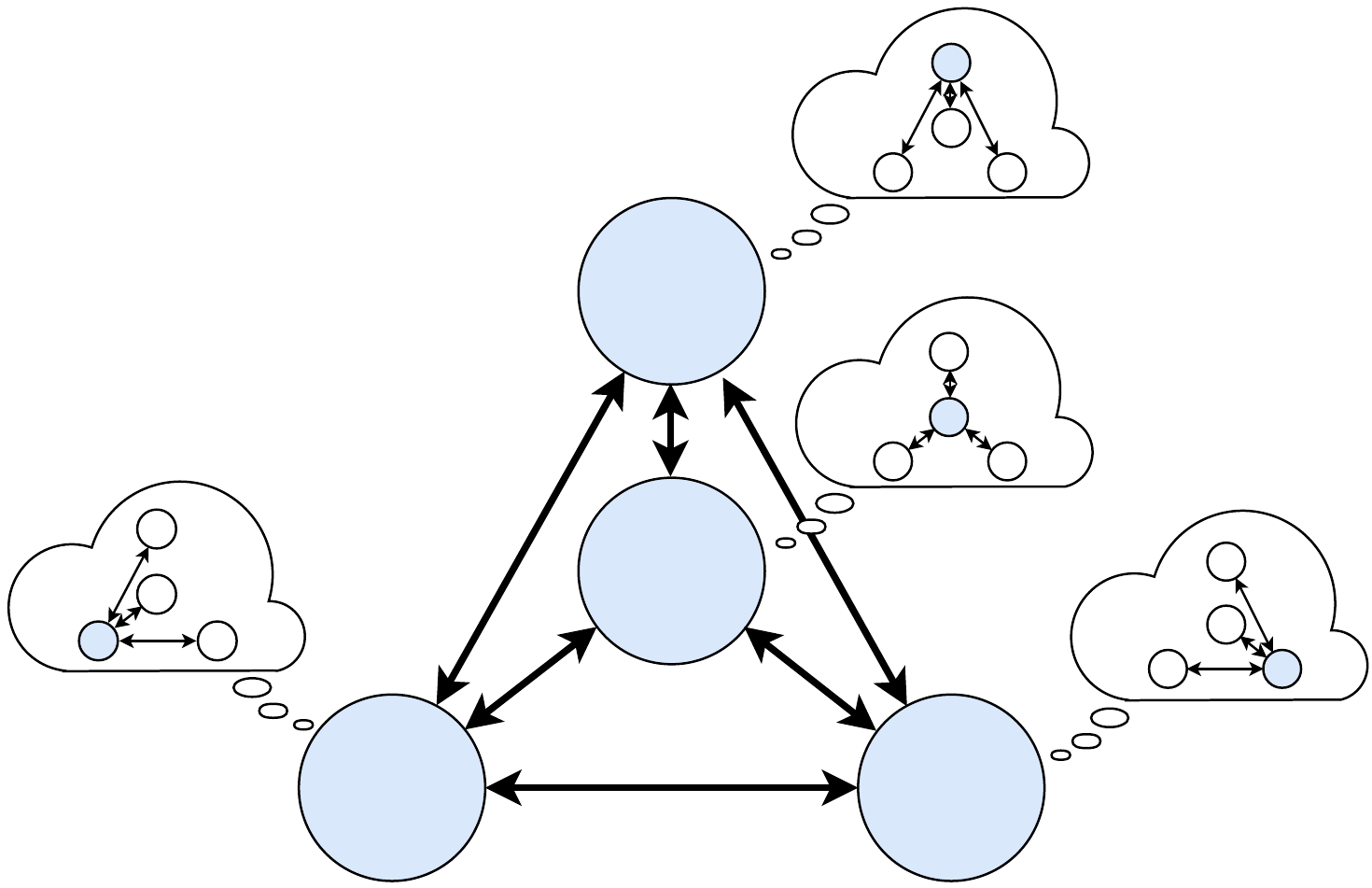}
    \caption{\label{fig:sharedworld}}
    \end{subfigure}
\caption{(a) Centralized, (b) distributed, and (c) shared-world communication architectures. In a centralized architecture, a single agent (shown in blue) collects relevant information from all agents, computes a plan for the entire system, and communicates the tasks or actions that each agent should perform to them. In a distributed architecture, agents only communicate with their neighbors. Finally, in a shared-world architecture, each agent executes an instance of a centralized algorithm; agents exchange relevant states and observations with all neighbors so as to ensure that the inputs to each agent's instance of the algorithm are the same.}
\label{fig:Centralized_distributed_SharedWorld}
\end{figure}

\revXXIa{
Algorithms for collective behavior can be centralized or distributed.

\textit{Centralized} algorithms require as input states and/or observations from all agents in the system; Their output is a collection of actions for each of the agents. 

Conversely, in \emph{distributed} algorithms, each agent executes an instance of the algorithm. The algorithm may prescribe the agent to exchange messages with other agents it can directly communicate with (the agent's \emph{neighbors}). The inputs to each agent's instance of the algorithm are restricted to be (i) agent's own state and (ii) messages received from the agent's neighbors.  The output of each agent's instance of the algorithm is a set of actions for the agent itself.

Centralized algorithms can be implemented in \textit{centralized} communication architectures. In such architectures, all agents  share their information with a central node, which executes the centralized algorithm and transmits a set of control actions to each agent (Figure \ref{fig:centralized}).

Distributed algorithms can be directly implemented on \emph{distributed} communication architectures, where agents can only share information with their neighbors (Figure \ref{fig:distributed}). 

\textit{Shared-world} architectures allow centralized algorithms to be executed on distributed communication architectures. In shared-world architectures, each agent executes a copy of the centralized algorithm (Figure \ref{fig:sharedworld}). Agents share relevant information about their state with every other agent (possibly through multi-hop communication) to ensure that everyone has access to the same inputs for the algorithm. Each agent  only performs the actions assigned to itself by the centralized algorithm, implicitly coordinating its actions with others. Shared-world implementations of centralized algorithms have been used  for multi-UAV tracking \cite{Ref:Campbell07}, planning for teams of autonomous maritime vehicles \cite{Ref:Elkins10,Ref:Wolf17,Ref:Sotzing10}, goal seeking \cite{Ref:Eich14,Ref:Leonard16}, and spacecraft formation control \cite{Ref:Beard01}.

Distributed algorithms can also be implemented in centralized communication architectures in a straightforward manner by simulating the message-passing process; for certain applications, executing this approach can lead to better scalability compared to a centralized algorithm (e.g. \cite{Ref:Boyd11}).
}

\newpage
\ifonecolumn
\processdelayedfloats
\fi

\section{Author Biography}

\ifonecolumn
Federico Rossi
\else
\begin{wrapfigure}{r}{0.15\textwidth}
   \centering
   \includegraphics[width=0.15\textwidth]{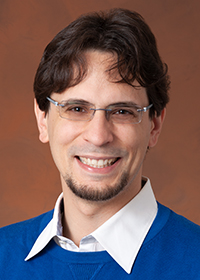}
\end{wrapfigure}
\textbf{Federico Rossi}
\fi
 is a Robotics Technologist at the Jet Propulsion Laboratory, California Institute of Technology.
He earned a Ph.D. in Aeronautics and Astronautics from Stanford University in 2018, a M.Sc. in Space Engineering from Politecnico di Milano and the Diploma from the Alta Scuola Politecnica in 2013.
His research focuses on optimal control and distributed decision-making in multi-agent robotic systems, with applications to robotic planetary exploration and coordination of fleets of self-driving vehicles for autonomous mobility-on-demand in urban environments.

\ifonecolumn
Saptarshi Bandyopadhyay
\else
\begin{wrapfigure}{r}{0.15\textwidth}
   \centering
   \includegraphics[width=0.15\textwidth]{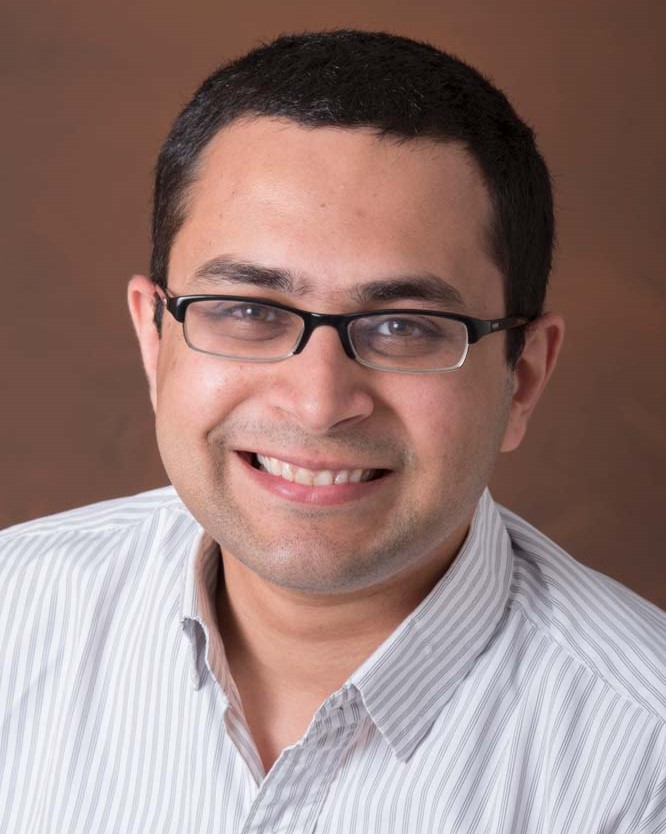}
\end{wrapfigure}
\textbf{Saptarshi Bandyopadhyay}
\fi
is currently a Robotics Technologist at the Jet Propulsion Laboratory, California Institute of Technology. He received his Ph.D. in Aerospace Engineering in January 2016 from the University of Illinois at Urbana-Champaign (UIUC). Saptarshi received his Dual Degree (B.Tech and M.Tech) in Aerospace Engineering in 2010 from the Indian Institute of Technology Bombay, India. At IIT Bombay, he co-founded and led the student satellite project. IIT Bombay's Pratham satellite was launched into low Earth orbit in September 2016. Saptarshi won the gold medal for India at the 9th International Astronomy Olympiad held in Ukraine in 2004. Saptarshi's research interests include aerospace systems, robotics, multi- agent systems and swarms, dynamics and controls, estimation theory, probability theory, and systems engineering.

\ifonecolumn
Michael T. Wolf
\else
\begin{wrapfigure}{r}{0.15\textwidth}
   \centering
   \includegraphics[width=0.15\textwidth]{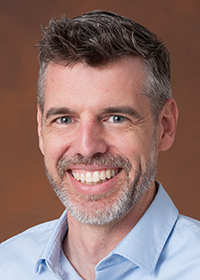}
\end{wrapfigure}
\textbf{Michael T. Wolf}
\fi
is a Principal Applied Scientist at Amazon Robotics AI. Prior to that, he was a Principal Research Technologist and the supervisor of the Maritime \& Multi-Agent Autonomy Group at the Jet Propulsion Laboratory, California Institute of Technology. His research efforts aim to improve practical, scalable approaches for autonomous mobility and multi-robot coordination, spanning work in perception, motion planning, task planning, and autonomy software architectures. Dr. Wolf received his Ph.D. and M.S. in Mechanical Engineering from Caltech and a B.S. in Mechanical Engineering from Stanford University.

\ifonecolumn
Marco Pavone
\else
\begin{wrapfigure}{r}{0.15\textwidth}
   \centering
   \includegraphics[width=0.15\textwidth]{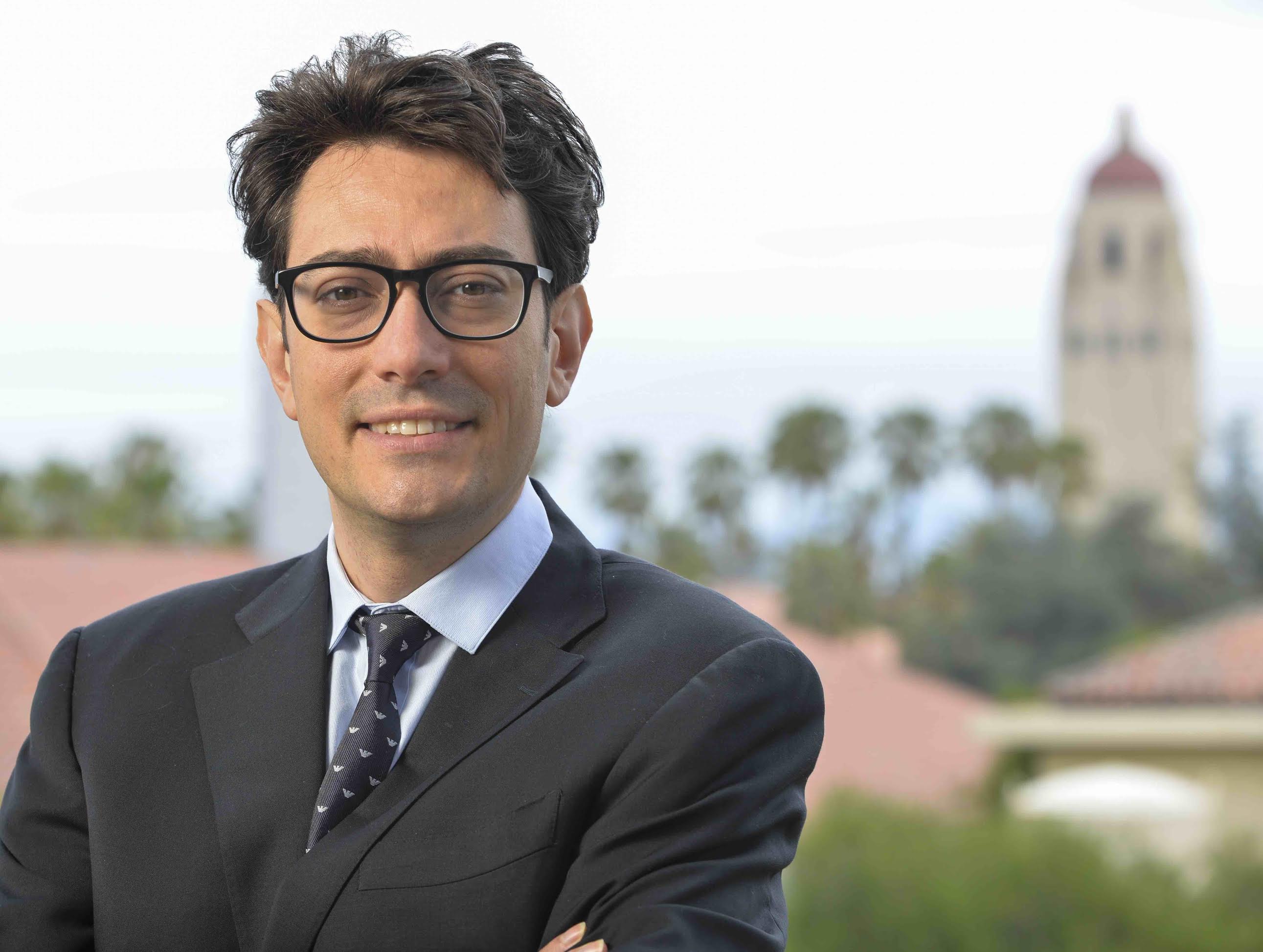}
\end{wrapfigure}
\textbf{Marco Pavone}
\fi
is an Associate Professor of Aeronautics and Astronautics at Stanford University, where he also holds courtesy appointments in the Department of Electrical Engineering, in the Institute for Computational and Mathematical Engineering, and in the Information Systems Laboratory. He is a Research Affiliate at the NASA Jet Propulsion Laboratory (JPL), California Institute of Technology. Before joining Stanford, he was a Research Technologist within the Robotics Section at JPL. He received a Ph.D. degree in Aeronautics and Astronautics from the Massachusetts Institute of Technology in 2010. Dr. Pavone's areas of expertise lie in the fields of controls and robotics.
Dr. Pavone is a recipient of an NSF CAREER Award, a NASA Early Career Faculty Award, a Hellman Faculty Scholar Award, and was named NASA NIAC Fellow in 2011. At JPL, Dr. Pavone worked on the end-to-end optimization of the mission architecture for the Mars sample return mission. He has designed control algorithms for formation flying that have been successfully tested on board the International Space Station.

\end{document}